\documentclass[%
 reprint,nofootinbib,
showpacs,preprintnumbers,
 amsmath,amssymb,
 aps,
 prd,
 longbibliography,
]{revtex4-1}

\pdfoutput=1

\usepackage{cancel}
\usepackage{mciteplus,slashed}
\usepackage{amssymb,graphicx,cancel,amsmath}
\usepackage{dcolumn}
\usepackage{bm}
\usepackage{subcaption}
\usepackage{hyperref}
\usepackage[capitalise]{cleveref}
\usepackage[T1]{fontenc}
\usepackage{booktabs}
\usepackage{float}
\usepackage{multirow}
\usepackage{physics}
\usepackage{siunitx}
\graphicspath{{Figures/}}

\newcommand{\be}{\begin{equation}}
\newcommand{\ee}{\end{equation}}
\newcommand{\ba}{\begin{align*}}
\newcommand{\ea}{\end{align*}}
\newcommand{\bpm}{\begin{pmatrix}}
\newcommand{\epm}{\end{pmatrix}}
\newcommand{\bea}{\begin{eqnarray}}
\newcommand{\eea}{\end{eqnarray}}

\newcommand{\benum}{\begin{enumerate}}
\newcommand{\eenum}{\end{enumerate}}
\newcommand{\bi}{\begin{itemize}}
\newcommand{\ei}{\end{itemize}}

\newcommand{\GeV}{~\mathrm{GeV}}
\newcommand{\TeV}{~\mathrm{TeV}}

\newcommand{\gsim}{\lower.7ex\hbox{$\;\stackrel{\textstyle>}{\sim}\;$}}
\newcommand{\lsim}{\lower.7ex\hbox{$\;\stackrel{\textstyle<}{\sim}\;$}}

\crefname{appsec}{Appendix}{Appendices}
\newcommand{\citer}[1]{Ref. \cite{#1}}

\begin{document}

\preprint{APS/123-QED}

\title{Probing new charged scalars with neutrino trident production}

\author{Gabriel Magill}
\email{gmagill@perimeterinstitute.ca}
\author{Ryan Plestid}
\email{plestird@mcmaster.ca}
\affiliation{Department of Physics and Astronomy, McMaster University, Hamilton, Ontario, Canada}
\affiliation{Perimeter Institute for Theoretical Physics, Waterloo, Ontario, Canada}
\date{\today}

\begin{abstract}
We investigate the possibility of using neutrino trident production to probe leptophilic charged scalars at future high intensity neutrino experiments. We show that under specific assumptions, this production process can provide competitive sensitivity for generic charged scalars as compared to common existing bounds. We also investigate how the recently proposed mixed-flavour production - where the two oppositely charged leptons in the final state need not be muon flavoured - can give a 20-50\% increase in sensitivity for certain configurations of new physics couplings as compared to traditional trident modes. We then categorize all renormalizable leptophilic scalar extensions based on their representation under $SU(2)\times U(1)$, and discuss the Higgs triplet and Zee-Babu models as explicit UV realizations. We find that the inclusion of additional doubly charged scalars and the need to reproduce neutrino masses make trident production uncompetitive with current bounds for these specific UV completions. Our work represents the first application of neutrino trident production to study charged scalars, and of mixed-flavour final states to study physics beyond the Standard Model. 
%
\end{abstract}

\pacs{12.15.Ji,12.60.Fr,13.15.+g,14.60.Pq}

\maketitle



\section{Introduction \& Motivation\label{sec:Intro}}
Neutrino oscillation experiments provide conclusive evidence that the Standard Model (SM) is incomplete. Many unresolved anomalies---the proton radius puzzle \cite{Pohl2013,Hill2017}, the anomalous magnetic moment of the muon \cite{Jegerlehner:2009ry, Bennett2006} and the LSND anomaly \cite{Aguilar2001}---can be interpreted as providing hints into beyond the SM (BSM) physics, especially for heavy leptons where constraints are typically weaker. Scalar extensions of the SM have been proposed as solutions to all of these anomalous measurements \cite{Babu1988,Babu2002,Lindner2016,Liu2016}. Currently, most constraints on the scalar sector come from low energy observables and high energy colliders \cite{Herrero-garcia2014,Dev2017}. In contrast, high intensity mid energy neutrino experiments have remained relatively uninvestigated. Consequently, new tools sensitive to interactions between scalars and neutrinos/heavy leptons provide a complimentary probe of beyond the SM (BSM) physics.

\begin{figure}[!t]
	\vspace{5pt}
	\centering
	\begin{center}
\includegraphics[width=0.9\linewidth]{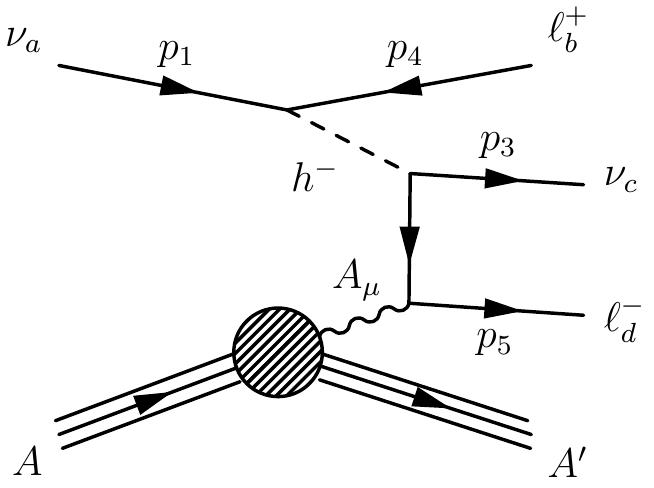}
	\end{center}
   \caption{Neutrino trident production of a charged Weyl lepton pair via a new charged scalar. There are three additional diagrams that can be obtained. The two charged leptons can be of different flavours. The connecting photon can interact with the nucleus (as shown above), or with individual nucleons.}
  \label{fig:scalartridentpicture}
  \end{figure}


Neutrino trident production (NTP) represents a natural candidate for studying couplings to an extended scalar sector given the successful application of NTP to models with an Abelian $Z'$ coupled to $L_\mu-L_\tau$ \cite{Altmannshofer2014}. Using data from the beam dump experiments CHARM-II and CCFR \cite{Geiregat1990,Mishra1991} the authors of \citer{Altmannshofer2014} were able to probe previously unexplored parameter space, including part of the favoured region for the resolution of the $(g-2)_\mu$ anomaly. As demonstrated in \citer{Magill2016}, the upcoming beam dump experiments SHiP and DUNE \cite{DUNECollaboration2015,SHiPCollaboration2015} are sensitive to many previously unmeasured neutrino trident channels which contain mixed-flavour leptons in the final state. With these exciting new prospects the possibility of NTP serving as a powerful probe of scalar extensions seems highly probable. Furthermore, given the mounting interest in precision neutrino physics, NTP may find applications at other future neutrino experiments, in particular Fermilab's Short-Baseline Neutrino program \cite{Chen2007}. 

NTP involves the creation of a lepton pair via  a high energy neutrino scattering coherently (diffractively) with a nucleus (nucleon) as shown in \cref{fig:scalartridentpicture}. This production mechanism is sub-dominant to charged-current (CC) scattering, in large part due to the extra $\alpha^2$ fine-structure suppression in its cross section; for $50 \GeV$ neutrinos scattering coherently on lead producing a $\mu^+\mu^-$ final state, we expect one trident event for every $10^5$ CC events \cite{Belusevic1988}. As discussed in \citer{Magill2016}, this scaling depends largely on the flavours of the final state lepton pair, with event rates being 40 times larger in the case of $e^+\mu^-$ production at DUNE. This is due to the absence of W-Z interference, and an infrared singularity in the phase space; the lower electron mass provides a log-enhanced cross section. Multi-flavour configurations were not observable in CCFR or CHARM-II due to difficulties in tagging electron final states. The potential to view these NTP processes at future experiments allows for a rich landscape of signals \cite{Lazvseth1971,Brown1972,Magill2016}. In particular, it lends itself to the study of off-diagonal lepton flavour couplings, and these appear naturally for new charged scalars. In this work, we study how these new mixed-flavour observables compare with existing probes of charged scalar theories that preserve the SM's $SU(2)\times U(1)$; we assume no additional fermion or vector content. The case of neutral scalars probed via the diagonal $\nu\mu^+\mu^-$ final state has been considered for a phenomenologically motivated Lagrangian in \citer{Ge:2017poy}. 

We find that charged scalars are best probed by NTP in the case of universal flavour diagonal couplings. For these configurations, we find that mixed-flavour trident final states can give a 20-50\% increase in sensitivity to BSM couplings as compared to the traditional $\nu\mu^+\mu^-$-trident channel, and consequently out-performing bounds from the anomalous magnetic moment of the muon. When considering explicit UV completions (such as a Higgs triplet), we characterize the experimental improvements one should make in order for bounds from NTP to be competitive. Additional neutral and doubly charged scalar particles often appear in the context of UV models reproducing neutrino oscillation data, and these can introduce new, and more stringent, constraints.

The rest of the article is organized as follows: In \cref{sec:trident}, we consider a general leptophilic charged scalar, how it contributes to trident, and its associated experimental backgrounds. For some benchmark choices in parameter space, we show the reach in sensitivity. In \cref{sec:scalarmodels}, we explain how our general model can arise by giving an exhaustive classification of all leptophilic, renormalizable and $SU(2)\times U(1)$ invariant scalar extensions. We discuss specific realizations of these classifications in the literature and in \cref{sec:constraints}, the phenomenological constraints surrounding them. We conclude with general remarks and potential applications in \cref{sec:conclusion}.


\section{Charged Scalar Mediated Trident Production}\label{sec:trident}
\subsection{Signal}
We consider a singly charged scalar coupling to the lepton doublets:
\begin{equation}
\mathcal{L}\supset |\partial_\mu h|^2 - m_h^2 |h|^2 + \sqrt{2}h_{ab}\nu^a\ell^bh +k_{ab}\ell^a\ell^bk + c.c. 
\label{eq:simplifiedNonUVmodel}
\end{equation}
The doubly charged scalar, which does not contribute to NTP, has been included to make connection with UV completions. The singly charged scalar contributes to NTP via diagrams like the one shown in \cref{fig:scalartridentpicture} and results in the amplitude shown in \cref{eq:zeebabu-matrix-element}. In the following, we use $x^\alpha$ and $y^\dagger_{\dot{\alpha}}$ to denote left- and right-handed initial states respectively, and $y^\alpha$ and $x^\dagger_{\dot{\alpha}}$ to denote right- and left-handed final states, following \cite{Dreiner:2008tw}. We assign the labels $\{1,2,3,4,5\}=\{\nu,\gamma,\nu',\ell^+,\ell^-\}$, and we use the mostly minus metric $\eta_{\mu\nu}=\{1,-1,-1,-1\}$.

In the context of the Equivalent Photon Approximation \cite{Budnev1975,Belusevic1988}, the matrix element for $\gamma\nu_a\rightarrow \ell^+_b \nu_c\ell^-_d$ can be summarized succinctly as 
\begin{equation}
  \begin{split}
    \mathcal{M}_{h}&=-\frac{2 h_{ab}h_{cd}^*e}{(P_1-P_4)^2-m_h^2}\left[\frac{\mathcal{A}_{14}}{q_d^2-m_d^2}+\frac{\mathcal{A}_{35}}{q_b^2-m_b^2}\right]\\
    \mathcal{A}_{14}&=\left\{\left(x_1 y_4\right)\left(x_3^\dagger\overline{\slashed{q}}_d \slashed{\epsilon}_2 x_5^\dagger\right) \
    +\left(x_1 y_4\right) \left(x_3^\dagger\overline{\slashed{\epsilon}}_2 y_5\right) m_d\right\}\\
    \mathcal{A}_{35}&=\left\{\left(x_1 \slashed{q}_b \overline{\slashed{\epsilon}}_2  y_4\right) \left(x_3^\dagger x_5^\dagger\right) \
    +\left(x_1\slashed{\epsilon}_2 x^\dagger_4\right)\left(x_3^\dagger x^\dagger_5\right) m_b\right\}\\
    q_b&\equiv P_2-P_4 ~~~~ q_d\equiv P_2-P_5 ~~~~\slashed{\ell}=\ell\cdot\sigma ~~~\overline{\slashed{\ell}}=\ell\cdot\overline{\sigma}.
\end{split}
\label{eq:zeebabu-matrix-element}
\end{equation}
where $\{a,b,c,d\}$ label lepton generations. 

The above matrix element contains contributions from four different diagrams. Two contain mass insertions appearing in the second terms of $\mathcal{A}_{14}$ and $\mathcal{A}_{35}$. The two amplitudes correspond to the photon interacting with either the negatively or positively charged lepton. The following identities 
\begin{equation}
u=(x,y^\dagger)^T~~v=(y,x^\dagger)^T ~~ \overline{u}=(y,x^\dagger)~~\overline{v}=(x,y^\dagger)
\label{eq:weyl-to-dirac}
\end{equation}
can be used to re-write the amplitudes in terms of the Dirac spinors 
\begin{equation}
  \begin{split}
    \mathcal{A}_{14}&=\overline{v}_1 P_L v_4
    \overline{u}_3 P_R \left(\slashed{q}_d + m_d\right)\
    \slashed{\epsilon}_2 v_5\\ 
    \mathcal{A}_{35}&=\overline{v}_1 P_L\left(\slashed{q}_b + m_b\right)\
    \slashed{\epsilon}_2 v_4 \overline{u}_3 P_R  v_5.\\
  \end{split}
  \label{eq:diracME}
\end{equation}
As a check of our calculations, we used the symbolic manipulation language FORM \cite{Manipulation} and compared our results to \cite{Lazvseth1971}. LEP searches rule out charged Higgs for $m_{h^\pm}\lesssim 100\GeV$ on general grounds based solely on its electromagnetic interactions with the photon and $Z$ boson \cite{Aleph2013} and so we have ignored the four-momentum in the scalar's propagator. The full cross section is obtained from \cref{eq:zeebabu-matrix-element} by 
\begin{equation}
  \sigma_{N\nu}=\frac{Z^2\alpha}{\pi}
  \int_{m_{jk}^2}^S\frac{ \mathrm{d}s }{s}\sigma_{\gamma\nu}(s)
  \int_{(s/2E_\nu)^2}^\infty \frac{\mathrm{d}Q^2}{Q^2}F^2(Q^2),
  \label{eq:Three-Body}
\end{equation}
where $F(Q^2)$ above is the Woods-Saxon form factor \cite{Magill2016,Jentschura2009}. 

For generic NTP final states the SM and BSM contributions can both be treated as real. The sign of the interference will be dictated by the symmetry or anti-symmetry of the couplings in \cref{eq:zeebabu-matrix-element}, as well as the relative sign of the SM contribution. For a given NTP process, the presence of $Z$ and/or $W$ vector mediators will induce an axial ($C_A$) and vector ($C_V$) coupling, upon which the matrix element depends linearly \cite{Magill2016}. If the SM mediators are both $W$ and $Z$ bosons ($C_{V,A}>0$), we find a positive relative sign. When the mediator is only a $Z$ boson ($C_{V,A}<0$), we get a negative sign. When the mediator is only a $W$ boson ($C_{V,A}=1$), we find a positive sign for $m_+>m_-$ and a negative one when $m_+<m_-$; this effect is related to subtle helicity properties \cite{Magill2016}. For anti-symmetric couplings $h_{e\mu}=-h_{\mu e}$ the new physics part of the matrix element will carry an additional negative sign, while for the symmetric case ($h_{e\mu}=h_{\mu e}$), there will be a positive sign. The final results for the sign of the interference terms are shown in \cref{sec:projectedsensitivities}. For symmetric (anti-symmetric) couplings, we have mostly constructive (destructive) interference.

\subsection{Search Strategy and Backgrounds}
Many flavour combinations for the incoming neutrino, outgoing neutrino, and charged leptons are possible. In deciding which reaction channel is ideally suited to one's purposes, two strategies should be considered. First a channel with a relatively high SM contribution could be chosen, allowing for interference effects, which will be dominant in the limit of small coupling\footnote{This interference is not sensitive to the phases of the couplings, which can be expected on general grounds related to the arbitrary definitions of phases in $h_{ab}$ \cite{Nebot2007}.}. Neutrino beams are predominantly composed of $\nu_\mu$ and so, in considering interference driven signals, we will typically consider incident $\nu_\mu$. Phase space considerations cause NTP rates to favour lighter lepton masses \cite{Magill2016} and so we focus our analysis on final states with at least one electron, or positron. When considering the older experiments CCFR and CHARM-II we consider their reported observations of $\mu^+\mu^-$ production.

A complementary approach is to consider a production channel that is closed in the SM, but open in the case of new physics. To ensure low backgrounds, one needs to be able to control the flux of incident (anti-)neutrinos. To see this consider $\nu_\mu \rightarrow e^-\mu^+\nu$ which is SM forbidden, but possible in the presence of BSM scalars. If, however, the beam was contaminated with $\overline{\nu}_\mu$ then the SM allowed $\overline{\nu}_\mu\rightarrow \mu^+e^-\overline{\nu}_e$ would present a substantial background. DUNE has the capability to eliminate contamination with its neutrino horn. In contrast, SHiP has a much more complicated incident flux profile and cannot separate the neutrino and anti-neutrino fluxes.

The sensitivities we present in this paper are based on future experiments measuring rates consistent with the irreducible SM coherent NTP backgrounds. The details of the procedure are outlined in \cref{sec:projectedsensitivities}. A full simulation would have to be performed by the collaborations prior to their analysis, but we believe our analysis provides a good approximation. For simplicity, we will focus the discussion at SHiP, however DUNE is also well equipped to tackle the same backgrounds. 

The SHiP tau neutrino detector, modelled after the OPERA experiment \cite{Agafonova2015}, is based on Emulsion Cloud Chambers (ECC) technology. The ECC is composed of a series of thin films interleaved with lead plates, followed by a muon spectrometer. A qualitatively similar setup to this was used in the CCFR experiment \cite{Mishra1991, KING1991254}, which featured iron plates interleaved with liquid scintillators and drift chambers. The use of fine emulsion film layers will provide SHiP with more accurate track ID capabilities as compared to CCFR. That said, CCFR was able to observe a $\mu^+\mu^-$ trident rate of 37 events given a theoretical SM prediction of 45 events. They isolated their signal by collecting $\mu^-$ and $\mu^+$ events and imposing cuts on the energy, angles, total invariant mass, hadronic activity, and vertex resolution. SHiP can implement similar cuts, however one caveat is that CCFR was dealing with much larger incoming neutrino average energies ($\sim 160\GeV$) as compared with SHiP and DUNE providing it with an enhanced signal \cite{Magill2016,Belusevic1988}. Since the bulk of these trident events are expected to come from SM processes, the kinematics of the outgoing pair of charged leptons is well captured in \cite{Lazvseth1971, Brown1972}. \\

Consider a mixed-flavour $\ell_b^+\ell_d^-$ lepton pair search with a hadronic veto. Final $e^+e^-$ states can arise from resonant $\pi^0$ production followed by a Dalitz decay where one of the photons is lost. For $\mu^-\mu^+$ final states, the dominant backgrounds will be from $\nu_\mu A\rightarrow \mu^- Y X$, where $Y$ represents either a  pion, kaon, charm- or $D$-meson which decays to a final state involving $\mu^+$ \cite{Adams:1999mn}, as seen by NuTeV. Production of vector meson final states is also likely, but these can be distinguished from NTP since they deposit more hadronic energy and lead to a larger invariant mass for the lepton pair. The decay length of pions is on the order of a few meters, and therefore these backgrounds could also contribute to $\mu^-e^+$ mixed-flavour final states if the meson fakes a charged lepton before decaying. The fake rate suppression at SHiP is very competitive. In particular, for electron ID efficiencies greater than $80\%$, the pion contamination rate is roughly $\eta_{\pi\rightarrow e}\le 0.5\%$ \cite{Buonaura2015}. In the SHiP detector at the end of the decay chamber, pion contaminations of $\eta_{\pi\rightarrow \mu}=0.1\%$ can be achieved for muon identification efficiencies of roughly 1.
%
For $e^-$ and $\mu^+$ final states, it is difficult to imagine how this would be produced outside of NTP. One possibility is coherent pion production from a $\bar{\nu}_e$ incoming state and a negatively charged pion. This background is expected to be small for a number of reasons, owing to the differences in the $\ell^+$ and $\ell^-$ energy spectrum, the much smaller lifetime flux of $\bar{\nu}_e$ at SHiP. Combinatorial backgrounds where one observes an electron and an anti-muon from two unrelated processes could be eliminated by the micron vertex resolution available at both SHiP and DUNE. 

\subsection{Model Independent Results}

\begin{figure}[!ht]
\includegraphics[width=\linewidth]
{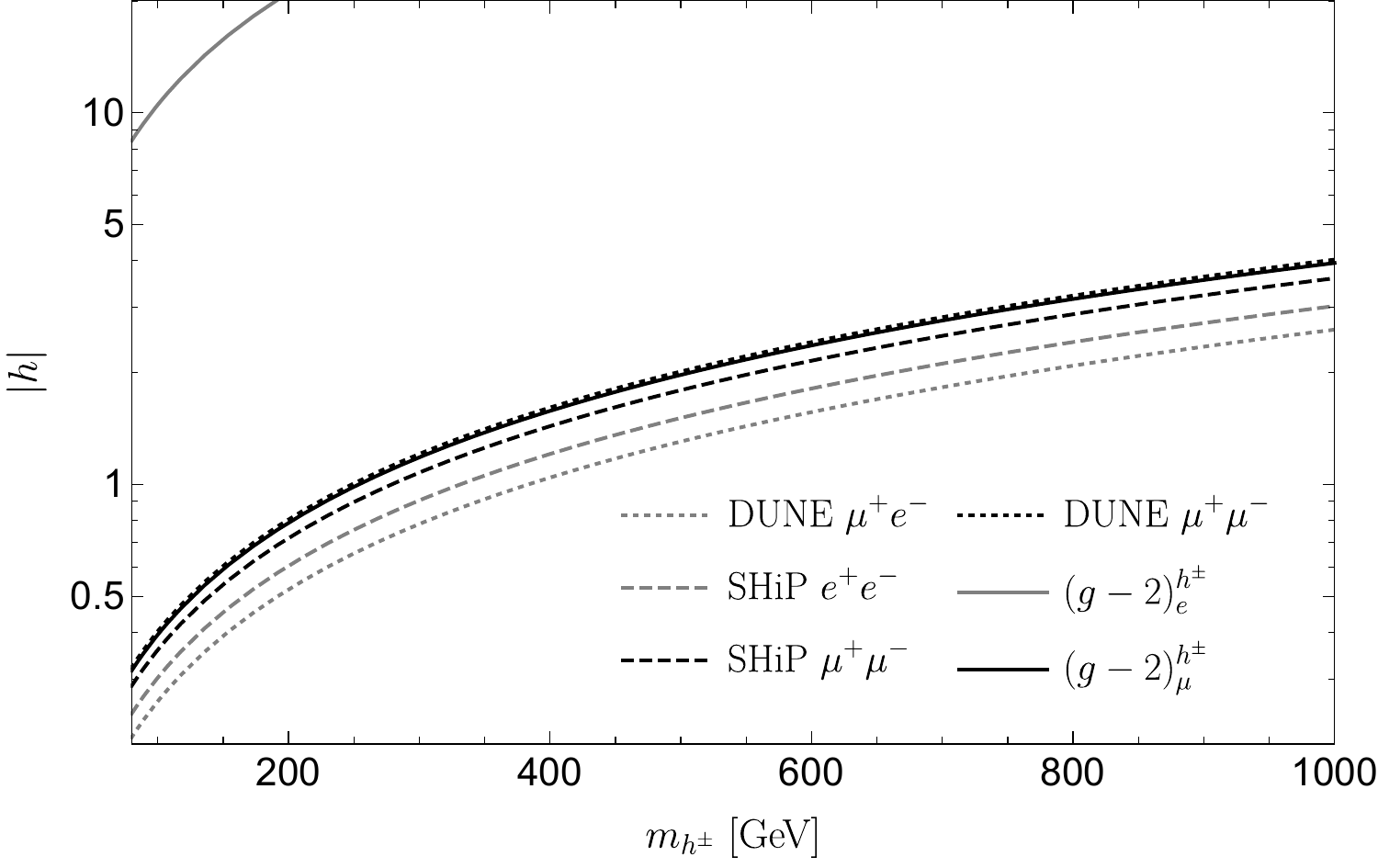}
\caption{\label{fig:TripletGeneraltDiagonal} Projected 90\% C.L. sensitivities at DUNE and SHiP for a given pair of final state oppositely charged leptons, and competing constraints when allowing only $h_{ee}=h_{\mu\mu}=h_{\tau\tau}\ne 0$.}
\label{fig:SHiP-DUNE-triplet-tmumuttautau}
\end{figure}

In this section, we illustrate the sensitivity of mixed-flavour NTP to charged scalars. We also highlight how certain flavour configurations  precluded in the SM give superior sensitivity to existing constraints. As an illustrative example, we consider the model described by \cref{eq:simplifiedNonUVmodel} and assume that $h_{aa}=|h|$, and $h_{ab}=0$ for $a\ne b$. As will be eventually discussed in \cref{sec:constraintsREALOG}, most of the strong existing constraints commonly considered for these types of models \cite{Herrero-garcia2014,Dev2017} drop out and NTP provides the dominant constraint, outperforming the $(g-2)$ for the muon. Our results are shown in \cref{fig:SHiP-DUNE-triplet-tmumuttautau}.

We have forecasted the SM backgrounds at SHiP and DUNE using the rates presented in \cite{Magill2016}. The best performing mode is the $\mu^+e^-$ channel at DUNE. A priori, the irreducible backgrounds to this process are $\overline{\nu}_\mu\rightarrow \mu^+\overline{\nu}_e e^-$ and $\nu_e\rightarrow \mu^+\nu_\mu e^-$. However, DUNE will have the ability to run in neutrino and anti-neutrino mode independently. This, coupled to the fact that the $\nu_e$ luminosities are low at this experiment makes this channel a 0 irreducible background search. Hence, we can use this channel to investigate the interplay between 0 background and the lack of interference term in the cross section. We make the interesting observation that the mixed-flavour final states in both experiments provide stronger constraints than the $\mu^+\mu^-$ states, while probing a Yukawa diagonal theory. In going from the muon final states to more general final states, the sensitivities to $|h|$ at DUNE are improved by $50\%$ whereas at SHiP, they can be improved up to $20\%$. 

\begin{figure}[h]
\includegraphics[width=\linewidth]
{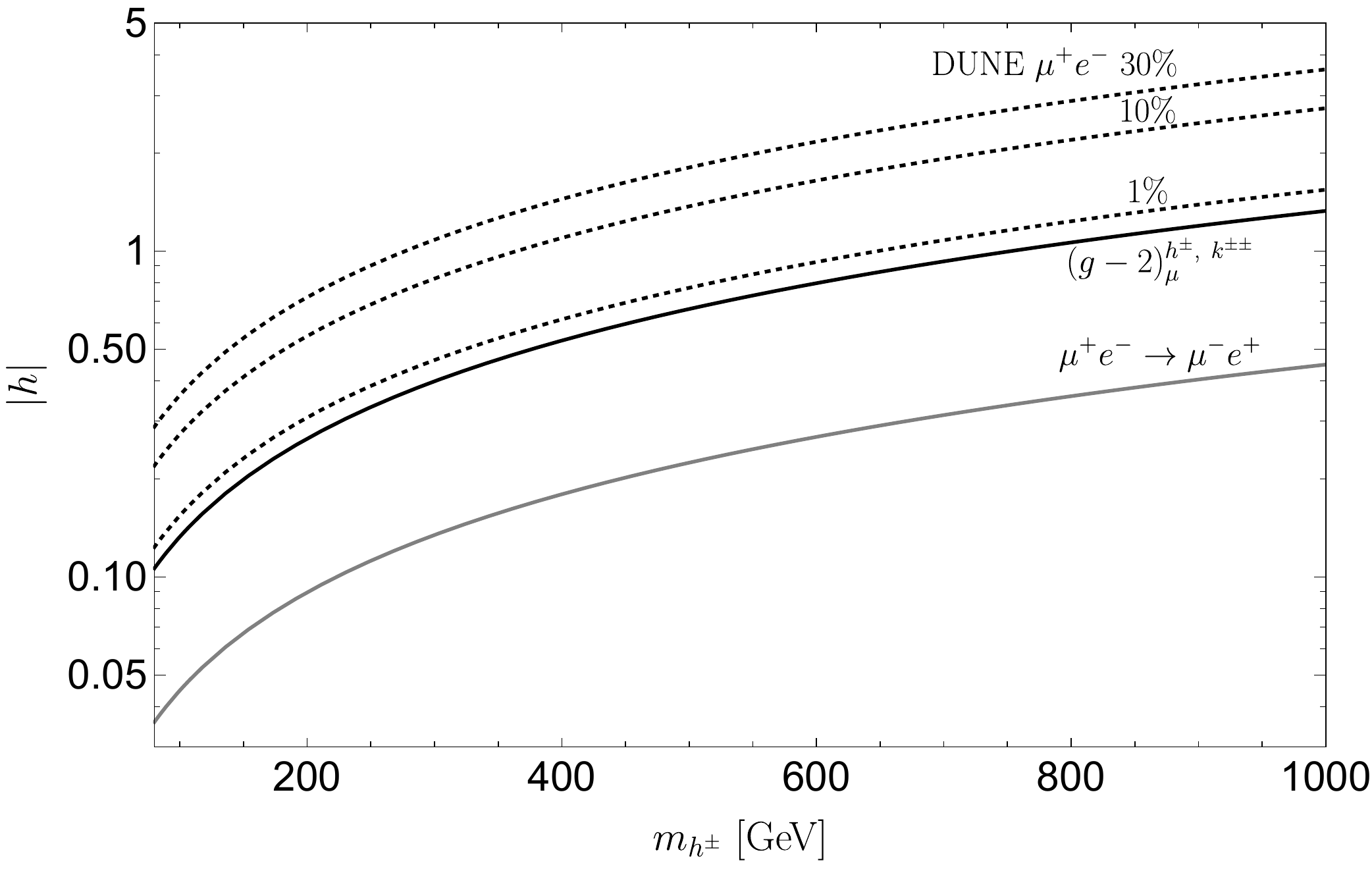}
\caption{\label{fig:TripletGeneraltDiagonalCrossSection} Projected sensitivities of NTP assuming the SM prediction at DUNE has been measured to various precisions measured as a percentage of the SM cross section. We compare this to other constraints which now include a doubly charged scalar.}
\end{figure}

We now show how NTP compares to other constraints when taking into account doubly charged scalars, assuming that $h_{ab}=k_{ab}$ in \cref{eq:zeebabu-matrix-element}. This is analogous to the HT model to be discussed later, without imposing the requirements of reproducing neutrino masses. The introduction of a doubly charged scalar $k^{\pm\pm}$ implies additional constraints from $\mu^+e^-\rightarrow \mu^-e^+$. A natural question to ask is what improvements in sensitivity are required to make trident competitive with these stronger constrains. We assume that one could measure the NTP cross section to within a given percentage of the SM cross section, for various benchmark precisions. These results are presented in \cref{fig:TripletGeneraltDiagonalCrossSection}. As a reference, the $10\%$ curve for DUNE's $\mu^+e^-$ channel corresponds roughly to the 90\% C.L. bounds shown in \cref{fig:TripletGeneraltDiagonal}. As can be seen, very high precision in the measured NTP cross section would be required to compete with the leading constraints on scalar couplings assuming $k^{\pm\pm}$.

\section{Extensions above the electroweak scale} \label{sec:scalarmodels}
We now illustrate how the phenomenological charged scalar model from \cref{sec:trident} can minimally arise while obeying all of the symmetries of the SM, with no additional fermion or vector matter content. The lepton sector's $SU(2)\times U(1)$ structure restricts possible scalar couplings that are relevant for NTP. The relevant leptonic fields are the $SU(2)$ doublets and singlets denoted by
\begin{equation}
\begin{array}{cc}
L_a^i=\begin{pmatrix} 
	\nu_a\\
    \ell_a
	\end{pmatrix},~~ \ell_a^c
\end{array}
\label{eq:2-fields}
\end{equation}
respectively, where $i$ labels the $SU(2)$ index, and  $a\in\{e,\mu,\tau\}$ labels the generations. All fields above are two-component left-handed spinors, with the spinor indices suppressed (i.e. $\ell_a^c=(\ell_a^c)_\alpha$). To couple these fermions to a scalar via a renormalizable interaction we can consider at most two lepton fields and one scalar. The possibilities are given in \cref{tab:class},
\begin{table}[h!]
\begin{tabular}{ccccc}
\toprule
Field & ~~$U(1)$ & $SU(2)$ & $\mathcal{L}_{\text{int}}$ & Couplings\\
\midrule
$\mathcal{S}$ & ~-2              & $1$ & $\mathcal{S}\ell^c_a\ell^c_b$                           & $s_{\{ab\}}$\\
$\mathcal{F}$ & ~~1              & $1$ & $\mathcal{F}~\epsilon_{ij}L_a^{[i}L_b^{j]}$       & $f_{[ab]}$\\
$\mathcal{D}$ & $-\tfrac{1}{2}$           & $2$ & $\mathcal{D}_{i} L_a^{i} \ell_b^c$     &      $d_{ab}$ \\
$\mathcal{T}$ & ~~1              & $3$ & $\mathcal{T}_{\{ij\}}L_a^{\{i}L_b^{j\}}$      & $t_{\{ab\}}$\\
\bottomrule
\end{tabular}
\caption{Classification of renormalizable lepton-scalar operators consistent with gauge invariance. The final column denotes the flavour symmetry ($\{ab\}$) or anti-symmetry ($[ab]$) due to the $SU(2)$ structure.}
\label{tab:class}
\end{table}
where the lowercase letters represent generational coupling matrices, and the capital script letters are the scalar fields. In the order shown in the table, these are the symmetric singlet, anti-symmetric singlet, doublet, and triplet models. The $s_{ab}$ and $t_{ab}$ couplings are symmetric in their indices. As for $f_{ab}$, the anti-symmetry under the $i\leftrightarrow j$ forces $f_{ab}$ to be anti-symmetric under $a\leftrightarrow b$. The couplings for $d_{ab}$ are unconstrained. A doubly charged scalar such as $S$ cannot contribute to NTP at tree-level, and we therefore focus on the fields $\mathcal{F}$, $\mathcal{D}$, and $\mathcal{T}$ for the purposes of NTP. The primary effects of the $SU(2)$ symmetry are to
\begin{itemize}
\item Enforce a relation between couplings of the neutral, singly, and doubly charged scalars. This occurs for the triplet case and introduces additional constraints with which NTP must compete.
\item Generate flavour symmetries in the couplings which can lead to constructive or destructive interference.
\end{itemize}

To discuss specific implementations of the $\mathcal{D}$, $\mathcal{T}$ and $\mathcal{F}$ classifications, we respectively consider the Two-Higgs-doublet model, the Higgs triplet (HT) model (also known as type-II seesaw), and the Zee-Babu (ZB) model. The full details of these models \cite{Nebot2007,Herrero-garcia2014,Dev2017,Cao2009} are discussed in \cref{sec:uvcompletions}. Here, we summarize the important features of the latter two theories. HT and ZB models both generate neutrino masses and feature a doubly charged scalar. In the ZB model, the couplings of leptons to the singly charged and doubly charged scalars are allowed to vary independently, whereas in the HT model, they are identical. In order to preserve the $SU(2)\times U(1)$ structure of the SM, the HT model contains in addition a neutral scalar which only couples to neutrinos. Without any extra model building, the neutral scalars considered in this paper cannot contribute to NTP in contrast to the models considered in \citer{Ge:2017poy}. To help make the connection with \cref{sec:trident}, we show important coupling relations in \cref{tab:coupl-rel}. 
\begin{table}[h]
\begin{tabular}{ccccc}
\toprule
Scalar Extension & $\sqrt{2}h_{ab}$ & $h^\pm$ & $k_{ab}$  & $k^{\pm\pm}$\\ 
\midrule
Zee-Babu & $ 2f_{[ab]}$ & $\mathcal{F}$ & $ s_{\{ab\}} $              & $\mathcal{S}$ \\
Type-II See-Saw                          & $\sqrt{2}t_{\{ab\}}$ &
$\Delta^{\pm}$ & $t_{\{ab\}}$             & $\Delta^{\pm\pm}$\\
\bottomrule
\end{tabular}
\caption{Relationships between type-II seesaw, Zee-Babu, and generic couplings $h_{ab}$ and $k_{ab}$. Curly (square) braces mark the couplings' (anti-)symmetry.
\label{tab:coupl-rel}}
\end{table}
%
%
%

\section{Explicit UV completions \label{sec:constraints}} 

\subsection{Singlet-scalars}\label{sec:uvresults}
\begin{figure}[h]
\begin{subfigure}[h]{.4\textwidth}
\includegraphics[width=\linewidth]
{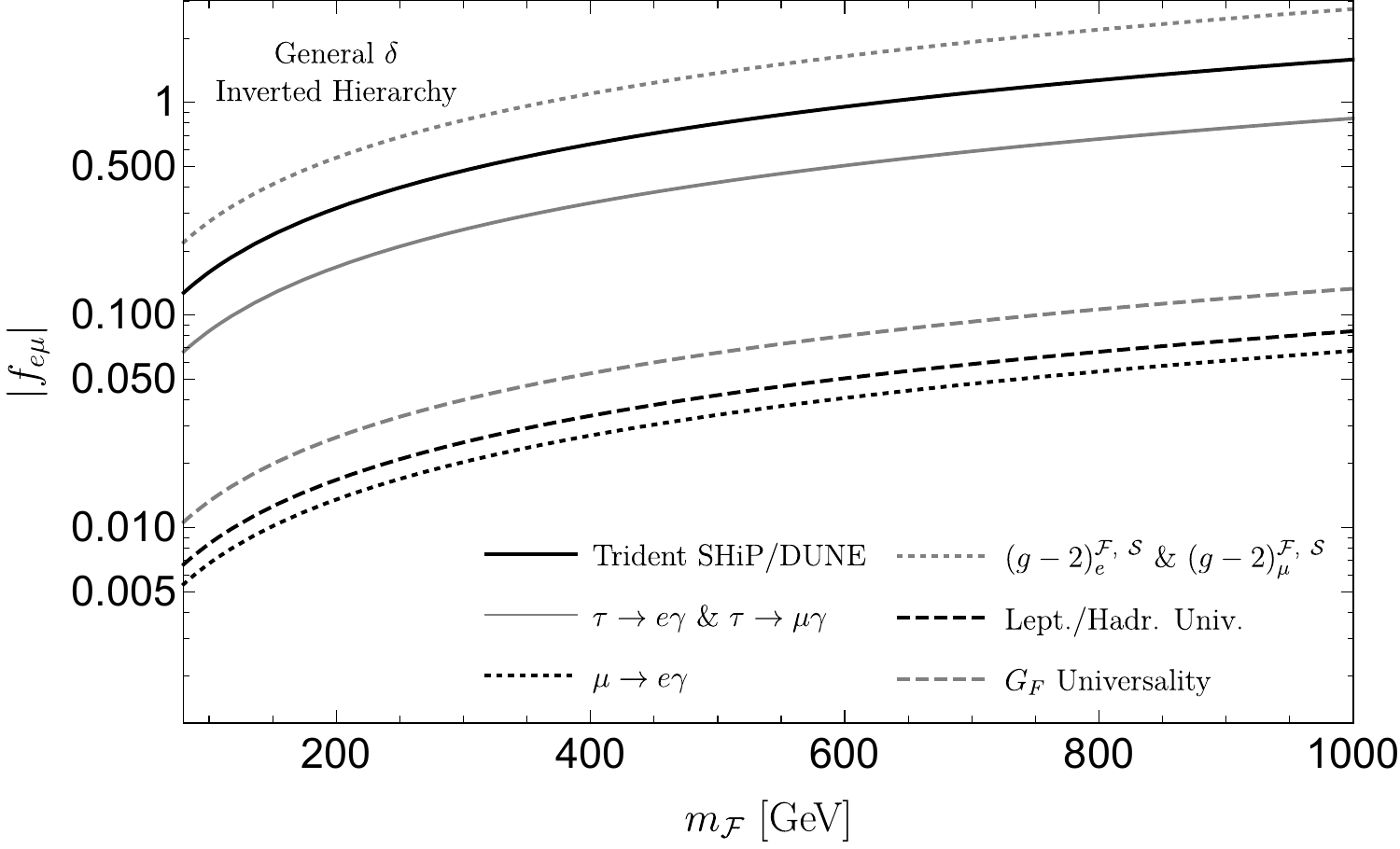}
\caption{\label{fig:ZeeBabuInvertedHiearchyDelta} Inverted Hierarchy\newline}
\end{subfigure}
\begin{subfigure}[h]{.4\textwidth}
\includegraphics[width=\linewidth]{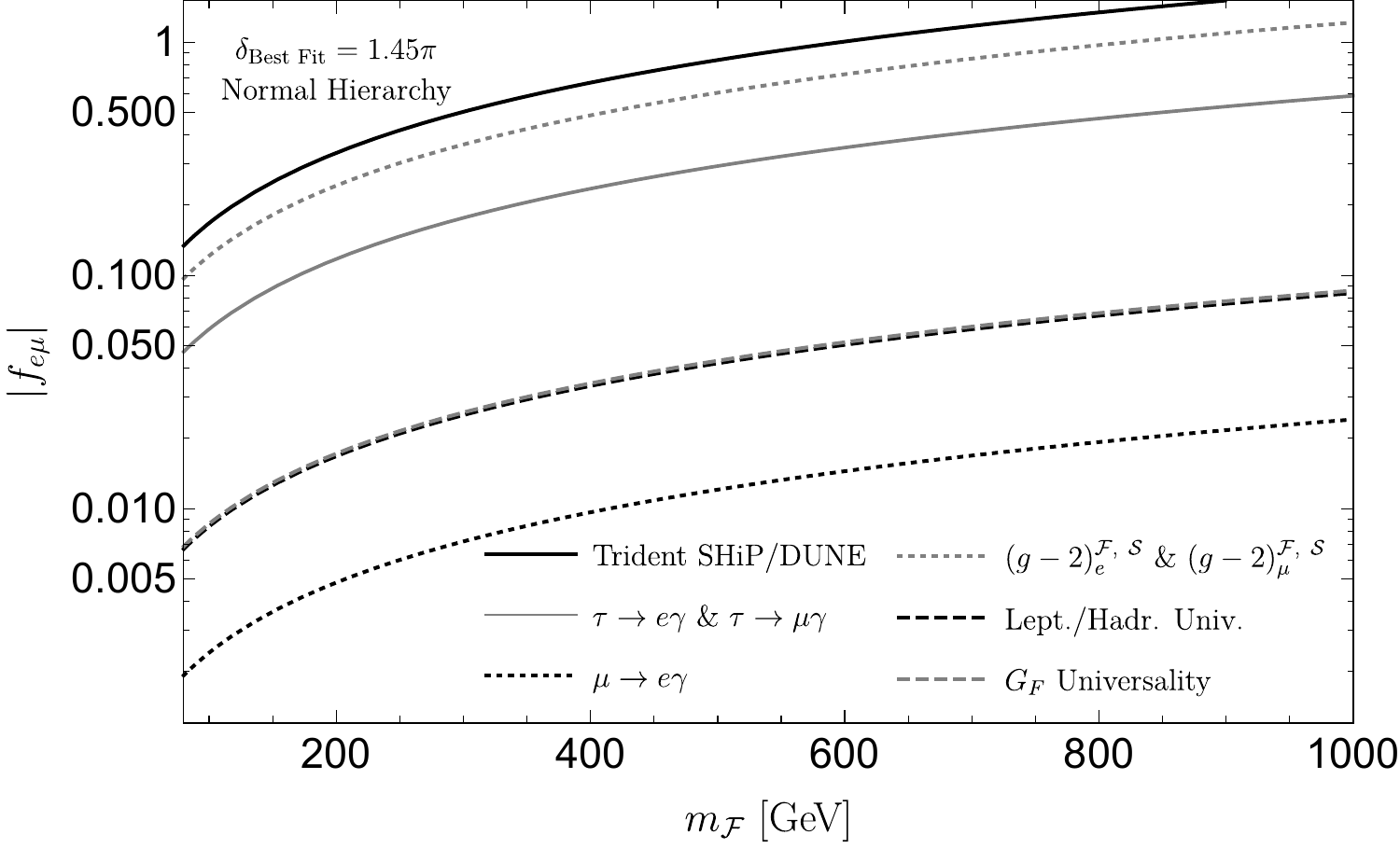}
\caption{\label{fig:ZeeBabuNormalHiearchyDeltaBest} Normal Hierarchy}
\end{subfigure}
\caption{Sensitivities for $|f_{e\mu}|$ assuming the Zee-Babu model generates neutrino masses.}
\end{figure}

We consider the ZB model to demonstrate the effects of negative interference and the requirements of reproducing neutrino textures. Using \cref{eq:2-texture5,eq:2-texture6,eq:2-texture7}, we express all of the $f_{ab}$ as a function of only $f_{e\mu}$ and PMNS matrix data \cite{Esteban2017}. Note that due to the vanishing $f_{aa}$ couplings of the ZB model, we are now probing non-diagonal couplings. We do this for both the normal and inverted hierarchies, and derive constraints on $|f_{e\mu}|$ as a function of $m_\mathcal{F}$ using the best performing mixed-flavour trident channels. For the normal hierarchy, we set the CP violating phase $\delta$ to its best fit value. For the inverted hierarchy, the dependence on $\delta$ factors out, and so the ZB model's contribution to NTP is independent of $\delta$. Our results are presented in \cref{fig:ZeeBabuNormalHiearchyDeltaBest,fig:ZeeBabuInvertedHiearchyDelta}.

\subsection{Triplet scalars: bounds from CCFR and CHARM-II}
The $\nu_\mu\rightarrow \nu_i \mu^+\mu^-$ final state was observed at the CCFR and CHARM-II experiments, and we can calculate experimental bounds on the triplet model using their data. Singlet scalars cannot be probed using this data due to the anti-symmetry of the couplings $f_{ab}$. 

CHARM-II had a neutrino beam of $\langle{E}_{\nu}\rangle\approx 20\GeV$ \cite{Geiregat1990,Altmannshofer2014} with a glass target ($Z=11$) and the CCFR collaboration had a neutrino beam of $\langle{E}_{\nu}\rangle\approx 160\GeV$ using an iron target ($Z=26$) \cite{Mishra1991,Altmannshofer2014}. The two experiments measured production cross sections of \cite{Altmannshofer2014}
\begin{align}
\begin{split}
\sigma_\text{CHARM-II}/\sigma_\text{SM} &= 1.58\pm 0.57\\
\sigma_\text{CCFR}/\sigma_\text{SM} &= 0.82\pm 0.28.
\end{split}
\end{align}
Using CCFR as an example, we set bounds by demanding 
\be
\sigma_{\text{SM}+\text{Triplet}}\le \sigma_\text{SM}(0.82 + 1.64\times 0.28),
\ee
where 1.64 standard deviations encompasses $90\%$ of a Gaussian likelihood function. For $m_\mathcal{T}$ in units of $\TeV$, 
\be
\left|t_{\mu \mu }\right|^2 \left[\frac{26.38}{m_\mathcal{T}^2}+1.59 \frac{\left|t_{\text{e$\mu $}}\right|{}^2+\left|t_{\mu \mu }\right|{}^2+\left|t_{\tau \mu }\right|{}^2}{m_\mathcal{T}^4}\right]\leq 691.36
   \label{eq:charmiiFullConstraint}
\ee
for CHARM-II, and 
\be
\left|t_{\mu \mu }\right|{}^2 \left[\frac{34.87}{m_\mathcal{T}^2}+1.97\frac{ \left|t_{\text{e$\mu $}}\right|{}^2+\left|t_{\mu \mu }\right|{}^2+   \left|t_{\tau \mu }\right|{}^2}{m_\mathcal{T}^4}\right]\leq 168.07  
   \label{eq:charmiiFullConstraint}
\ee
for CCFR. Assuming $|t_{ab}|=\abs{t}$, at $90\%$ C.L. the two collaborations impose the following constraints
\begin{equation}
\begin{split}
\abs{t}&\le 3.10\left(\frac{m_\mathcal{T}}{\TeV}\right) ~~~~~\text{CHARM-II}\\
\abs{t}&\le 1.77\left(\frac{m_\mathcal{T}}{\TeV}\right)~~~~~\text{CCFR}.
\end{split}
\end{equation}
The stronger bounds from CCFR are a result of the fact that this experiment saw a deficit of events in comparison to the SM prediction and so the upper-bound at $90\%$ C.L. is lower than CHARM-II.
%

\section{Conclusions \& Outlook}
\label{sec:conclusion}
We have investigated NTP as a tool for studying scalar extensions of the SM. We have considered $SU(2)$ singlet, doublet, and triplet charged scalar extensions that couple to leptons, and concluded triplet and singlet scalar can contribute appreciably to NTP. 

In the case of triplet extensions we have found that NTP can serve as a complementary probe of the scalar sector at future experiments. In particular, for specific choices of model parameters in which LFV bounds vanish, NTP provides greater sensitivity than measurements of the anomalous magnetic moment. We found NTP to provide comparable sensitivity for charged singlet scalars and previous $Z'$ models \cite{Altmannshofer2014} in phenomenologically allowed mass ranges, despite their very different interaction nature. These prospects could be improved as the intensity frontier is pushed forward, and NTP may prove to be a valuable tool in the future. For generic choices of parameters, it is unlikely that NTP can compete with strong LFV constraints.

We have considered both the upcoming experiments SHiP and DUNE. The advantage of DUNE is its ability to isolate beams of $\nu_\mu$ and $\overline{\nu}_\mu$ with high purity by using a magnetic horn. We have shown that this enables us to remove the irreducible background for certain processes, namely $\nu_\mu\rightarrow \mu^+e^-\nu_i$, which has no SM contribution and is a viable production process in triplet models. This has the advantage of providing a clean signal, but results in a sensitivity that scales as $|h|^4$, in contrast to interference effects which can dominate for small coupling and scale as $|h|^2$. The lack of interference with the SM in these particular modes hinders sensitivity. For other channels, the relative phase between the SM and new physics contribution was found to be highly dependent on initial states, which had a tendency to cause destructive (constructive) interference in singlet (triplet) mediated NTP cross sections as can be seen in \cref{sec:projectedsensitivities}. 

The advantages provided by DUNE's nearly mono-flavour beam must be balanced against its relatively low-Z detector (argon $Z=18$) as compared to SHiP (lead $Z=82$). Additionally DUNE uses a lower energy beam ($\langle E_\nu\rangle=5 \text{GeV}$ vs $\langle E_\nu\rangle=20 \text{GeV}$) but compensates for this via a higher number of protons on target. In contrast to DUNE, SHiP's future lead based detector provides an ideal setting to take advantage of the $Z^2$ coherent enhancement however the lack of a neutrino horn, and the mutli-flavour nature of the neutrino beam suggests that searches at SHiP will have higher SM irreducible backgrounds. 

Lastly, we have investigated representative UV models leading to the generic scalar extensions discussed above. In the ZB and HT models, extra particles and relations between couplings arise if the scalar sector is expected to produce empirically viable neutrino textures. The added constraints due to tree-level LFV decays mediated by the doubly charged scalar and from the LHC are especially strong, and in some sense NTP is less important.

The influence of final states on the phase of the SM contribution may be of interest in future applications of NTP to new physics. This dependence is dictated not only by the flavour combinations in the initial and final states, but also the relative sizes of the charged lepton masses. This final feature is a consequence of the chiral structure of the weak interaction \cite{Magill2016}. The influence of these relative phases would be easy to miss and will play a crucial role in any future work that relies on interference with the SM. Although we have considered charged scalars which are already very constrained, we expect many of the qualitative features present in our analysis to be applicable to broader classes of model. In particular the unique ability of mixed-flavour final states to control the presence or absence of constructive interference. Finally, we were able to identify final mixed-flavour states with no SM counterparts, thus removing irreducible backgrounds. Our results expand the reach of future neutrino experiments---such as DUNE, SHiP, and SBN---to physics beyond their main research programme, both within and beyond the SM.


\section*{Acknowledgements}
We are very grateful to Itay Yavin and Maxim Pospelov for their continued guidance and for suggesting mixed-flavour trident production and its potential applicability to scalar models in future intensity frontier experiments. We thank Brian Shuve and Wolfgang Altmannshofer for feedback on the manuscript, as well as Cliff Burgess, Richard Hill, Stefania Gori, Chien-Yi Chen and Sarah Dawson for useful discussions. This research was supported in part by Perimeter Institute for Theoretical Physics. Research at Perimeter Institute is supported by the Government of Canada through the Department of Innovation, Science and Economic Development and by the Province of Ontario through the Ministry of Research and Innovation. This research was also supported by funds from the National Science and Engineering Research Council of Canada (NSERC), and the Ontario Graduate Scholarship (OGS) program.

\appendix
\crefalias{section}{appsec}


\section{Explicit UV Completions}\label{sec:uvcompletions}
In what follows we discuss popular implementations of each class of scalar models outlined above. The ZB model, used to radiatively generate neutrino masses, is a representative candidate for singlet scalars $\mathcal{F}$ (see \cref{tab:class}). Two-Higgs-doublet model (2HDM) have been considered extensively in the literature as an implementation of doublets $\mathcal{D}$ and we discuss neutrino trident production's ability to probe their couplings below. Finally the type-II seesaw mechanism (as known as HT models) for the generation of neutrino masses is discussed as the canonical example of a triplet model $\mathcal{T}$. 

\subsection{Two-Higgs-doublet models}
2HDMs have been extensively studied \cite{Abe2015,Cao2009,Branco:2011iw}. In most implementations of a 2HDM there will be mixing between the new BSM and SM Higgs doublets. This suggests that for the model to be technically natural couplings between the BSM charged doublet and leptons should be mass-weighted to incorporate the influence of the mass-weighted SM Higgs field. In the SM, rates of NTP are log-enhanced by infrared phase space effects which are controlled by the small masses of the charged leptons. If one were to consider NTP mediated by the charged component of a doublet scalar extension, this small mass infrared enhancement would compete directly against the mass-weighted Yukawa coupling suppression. Our explicit sensitivity calculations confirm that these competing effects make trident uncompetitive with existing constraints. We note that in the absence of mass-weighted couplings, NTP may be able to address this interesting region of parameter space, however this situation is technically unnatural due to radiative corrections from the Higgs boson---which induces corrections proportional to the SM Yukawa couplings---and would require a new physics mechanism to avoid fine-tuning.

\subsection{Zee-Babu model}
\begin{figure}[th]
	\centering
	\begin{center}
\includegraphics[width=0.8\linewidth]{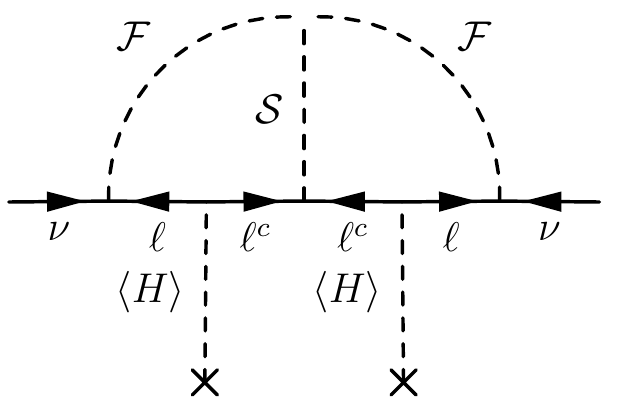}
	\end{center}
    \caption{Neutrino mass generation via the Zee-Babu model using two-component fermions with the direction of the arrows indicating chirality. 
      \cite{Dreiner:2008tw}. }
    \label{fig:tridentpicture}
  \end{figure}

A popular implementation of the scalar singlet model is the ZB model \cite{Cheng1980,Zee1985,Babu1988}. The model features a singly charged scalar $\mathcal{F}$ that couples to the leptonic doublets, and a doubly charged scalar $S$ which couples to the right-handed lepton singlets. The Yukawa sector of the Lagrangian can be written as:
\begin{equation}
\begin{split}
\mathcal{L}_{\text{ZB}}&\supset f_{ab}L_{a}^iL_{b}^j\epsilon_{ij}\mathcal{F} + s_{ab}\ell_a^c \ell_b^c \mathcal{S}+ h.c. \\
&=2f_{ab} \nu_a\ell_b\mathcal{F}+s_{ab}\ell_a^c \ell_b^c \mathcal{S}+ h.c. 
\end{split}
\label{eq:zeebabulagrangian}
\end{equation} 
This model is typically considered in the context of radiatively generated neutrino masses. These first occur at two loops via diagrams such as the one shown in \cref{fig:tridentpicture}. Assuming the ZB model is fully responsible for the generation of neutrino masses, the mass matrix $M$ can be expressed in terms of the ZB couplings $f_{ab}$ and the SM Yukawa couplings $Y$ via the relation $M\propto f Y s Y^T f^T$.
  %
%
The anti-symmetric matrix $f$ has odd dimensions and therefore its determinant will vanish by Jacobi's theorem. Since the neutrino mass matrix $M$ contains $f$, its determinant will also vanish. This indicates that the smallest neutrino mass $m_1$ ($m_3$) will vanish in the case of the normal (inverted) hierarchy. The presence of a 0 mass mode \cite{Herrero-garcia2014,Herrero-garcia2014a} for the normal hierarchy implies
%
%
\begin{align}
\frac{f_{e\tau}}{f_{\mu\tau}}&=\tan\theta_{12} \frac{\cos\theta_{23} }{\cos\theta_{13}}+\tan\theta_{13}\sin\theta_{23}e^{-i\delta}  \label{eq:2-texture5}\\
\frac{f_{e\mu}}{f_{\mu\tau}}&=\tan\theta_{12} \frac{\sin\theta_{23}}{\cos\theta_{13}}-\tan{\theta_{13}}\cos\theta_{23}e^{-i\delta}  \label{eq:2-texture6}
\end{align}
and for the inverted hierarchy yields
\begin{equation}
\frac{f_{e\tau}}{f_{\mu\tau}}=-\frac{\sin\theta_{23}}{\tan\theta_{13}}e^{-i\delta} ,~\frac{f_{e\mu}}{f_{\mu\tau}}=~\frac{\cos\theta_{23}}{\tan\theta_{13}}e^{-i\delta}. 
\label{eq:2-texture7}
\end{equation}
These relations will be used in the results of \cref{sec:uvresults}, as they provide definite relations between the phases of the various couplings. A phase convention must be chosen, and a simple choice is $0\leq f_{e\tau}\in \mathbb{R}$. Inspecting \cref{eq:2-texture7} reveals that if the ZB model is responsible for the observed neutrino textures, and the hierarchy is determined to be inverted, then $0>f_{e\mu}\in\mathbb{R}$, while
$\textrm{Arg}~f_{\mu\tau}=\delta+\pi$. The case of the normal hierarchy is somewhat more involved, however two limits, namely $\delta=0$ and $\delta=\pi$ result in all the couplings being real and positive by virtue of $\cos\theta_{23}\approx \sin\theta_{23}$ and $\tan{\theta_{13}}\ll \tan{\theta_{12}}$.


\subsection{Type-II seesaw mechanism}\label{sec:typeiiseesaw}
One of the most popular triplet scalar extensions arises in the context of the seesaw mechanism for generating neutrino masses, specifically the so called type-II seesaw or Higgs triplet model \cite{FileviezPerez2008,Sugiyama2013,Das2016}. In this version, a triplet field with matrix representation  
\be
\mathcal{T} \equiv i\sigma_2\cdot \Delta  \equiv
-\frac{1}{\sqrt{2}}\begin{pmatrix}
   	-\Delta^{0} &  	\Delta^+			\\
    \Delta^{+} &	\sqrt{2}\Delta^{++}	
\end{pmatrix}
\label{eq:triplethiggsrepresentation}
\ee
is introduced into a symmetric lepton product via an interaction of the form $t_{ab}L_a^i\mathcal{T}_{ij}L_b^j$. After $\Delta^0$ acquires a VEV $\textrm{v}_T$, we generate neutrino mass terms of the form $\textrm{v}_T\nu_a\nu_bt_{ab}$. This model has been ruled out by measurements of the invisible width of the $Z$ boson at LEP \cite{Montero1999,Langacker:1226768,Das2016}. These bounds can be evaded by softly breaking the symmetry in the Lagrangian with the terms
\be
-m_H^2H^2+(\mu H^Ti\sigma_2\Delta^\dagger H+h.c.)+M_\Delta^2\Tr(\Delta^\dagger\Delta)
\ee
where $\mu$ can be small to approximately retain the global symmetry. Minimizing this with respect $\Delta^0$ and setting $\langle \Delta^0 \rangle = \textrm{v}_T$ yields the equation
\be
\textrm{v}_T=\frac{\textrm{v}_d^2\mu}{\sqrt{2}M_\Delta^2},
\ee
where $\textrm{v}_d$ is the SM Higgs' VEV. Since the neutrino masses are given by $\textrm{v}_T t_{ab}$, we can generate small masses in the limit where $M>\textrm{v}\equiv \sqrt{\textrm{v}_d^2+\textrm{v}_T^2}=246\GeV$. As relevant for NTP, we have the Lagrangian
\begin{equation}
\mathcal{L}_\text{HT}\supset
-t_{ab}\left(\ell^a\Delta^+\sqrt{2}\nu^b+\ell^a\Delta^{++}\ell^b-\frac{\nu^a\Delta^0\nu^b}{\sqrt{2}}\right)+h.c.
\label{eq:tripletflavourbasis}
\end{equation}
The flavour symmetry of $t_{ab}$ allows for flavour diagonal terms in the Lagrangian to be non-vanishing. This is in contrast to the singly charged couplings $f_{ab}$ in the ZB model. The off-diagonal flavour couplings with $\Delta^\pm$, can be related to the ZB couplings as shown in \cref{tab:coupl-rel}. The $\Delta^0$ and $\Delta^{\pm\pm}$ scalars do not contribute to the NTP amplitudes, and so the trident exclusions we obtain on $t_{ab}$ come only from $\Delta^\pm$ leptonic interactions. These must compete with other phenomenological considerations which can be mediated by the $\Delta^{\pm\pm}$ or $\Delta^0$ fields. The propagating degrees of freedom of the scalar sector can in principle be different than the fields specified above \cite{FileviezPerez2008}. However, $\rho$ parameter constraints imply that the triplet VEV is at least 2 orders of magnitude smaller than the Higgs VEV \cite{Langacker:1226768}. Therefore the mixing will be very small and we can think of $\Delta^\pm$ as being the physical mass eigenstate.

\section{Constraints}
\label{sec:constraintsREALOG}
The addition of charged scalars to the SM leads to a variety of phenomenological consequences. In this section we discuss relevant constraints on the couplings involving the singly charged scalar ($h_{ab}$) and the couplings involving the doubly charged scalar ($k_{ab}$). The latter coupling does not play a role in NTP at tree level, however in the case of a triplet extension bounds on $k_{ab}$ can be converted to constraints on $h_{ab}$ since the two coupling matrices are related to one another as shown in \cref{tab:coupl-rel}. Below we review existing probes of the scalar sector, which we will compare with projected sensitivities using NTP as presented in \cref{sec:uvresults}.

\subsection{Anomalous magnetic moment measurements}
Charged scalars can alter a particle's magnetic moment \cite{Wu2001,Dicus2001,Pires2001}. Additionally there is a long-standing discrepancy between the measured value of $(g-2)_\mu$ and the SM prediction \cite{Bennett2006}. As a result there is some ambiguity in the interpretation of this measurement as either a prediction of the BSM theory or as a contraint on its couplings. These bounds are the weakest of those presented in \cite{Herrero-garcia2014,Dev2017}. As showed in \cref{sec:trident} for certain configurations of parameter space, NTP was capable of exceeding the sensitivity provided by this class of measurements. This is not surprising given NTP's competitive reach in the context of $Z'$ models as outlined in \cite{Altmannshofer2014}.

\subsection{Relative decay rates for $\mu$ and $\tau$ leptons}
Another class of constraints can be obtained by using the relative size of various measured leptonic decay rates
\be
\frac{\Gamma[\tau\rightarrow e/\mu + \text{inv.}]}{\Gamma[\mu \rightarrow e+\text{inv.}]}~~~~~\frac{\Gamma[\tau \rightarrow \mu+\text{inv.}]}{\Gamma[\tau\rightarrow e  + \text{inv.}]}
\label{eq:universalityMeasurements}
\ee
where `inv.' denotes invisible products (typically neutrinos). 
Measuring these quantities \cite{Pich2014} effectively measures the deviation from unity of flavour ratios of weak couplings $g_W^a/g_W^b$ for various flavours $a$ and $b$. Models with charged scalars will generically contribute to $\tau$ decays and so, the measurements of \cref{eq:universalityMeasurements} can be translated as bounds  on $||h_{i\tau}|^2-|h_{ej}|^2|$ as a function of the mass of $h^\pm$ \cite{Herrero-garcia2014}. From the arguments of \citer{Herrero-garcia2014,Nebot2007}, a singly charged scalar would contribute to the decay $\mu\rightarrow e\nu\overline{\nu}$, but would not affect beta decay. Therefore by using data reported in \citer{Pich2014} they were able to constrain  $\abs{h_{e\mu}}^2$, by considering a singly charged scalar's contribution to muon decay and noting that only final states with $e\nu_\mu \overline{\nu}_e$ would interfere with the SM amplitude. 
The quoted constraint\footnote{The full set of constraints as applied to the ZB model can be found in Table II and III of \citer{Herrero-garcia2014}, and model independent constraints can be obtained by setting the doubly charged scalar's coupling to zero.} is 
$\abs{h_{e\mu}}^2<0.014 \qty(\tfrac{m_h}{\text{TeV}})^2$ \cite{Herrero-garcia2014} after accounting for the normalizations shown in \cref{tab:coupl-rel}.

\subsection{Loop-level LFV decays}
LFV decays of the form $\ell_j\rightarrow \ell_i \gamma$ provide another tool to probe $h_{ab}$. This decay mode in the SM is extremely suppressed, and the observation of this LFV process would constitute strong evidence for new physics. Of particular interest is the decay mode $\mu\rightarrow e\gamma$ which provides the most stringent constraints on any of the couplings \cite{TheMEG:2016wtm}.

\subsection{Tree-level LFV decays}

In the case of triplet extensions where $h_{ab}$ and $k_{ab}$ are necessarily related (as shown in \cref{tab:coupl-rel}) strong upper limits on certain decay modes \cite{Herrero-garcia2014}, such as $\mu^-\rightarrow e^+ e^- e^-$, already precludes the regions of parameter space trident is capable of probing. On some level these constraints may be evaded by choices related to the Majorana phases in the mass matrix \cite{Merle2006,Grimus2012}, however we have not included these subtleties in our analysis. For singlet scalar extensions $k_{ab}$ and $h_{ab}$ are independent and NTP does not need to compete with bounds related to tree-level LFV decays.

\subsection{Implications of the LHC}
When including doubly charged scalars, LHC constraints become very strong. There are analyses by both CMS and ATLAS \cite{Chatrchyan:2012ya,ATLAS:2012hi} on doubly charged scalars decaying to same sign di-leptons which impose a model independent bound on the scalar mass of $200-400\GeV$. A recast \cite{delAguila:2013mia} of those LHC searches extended the constraints on the mass by an additional $100\GeV$ by explicitly requiring a total non-zero lepton number in the final state (by considering final states of same-sign dileptons and gauge bosons). In \citer{Das2016}, the authors showed that $h\rightarrow \gamma \gamma$ measurements at the LHC, the oblique $T$ parameter limits and exclusions from LEP implies a lower bound on $m_{h^\pm}$ as a function of the triplet VEV $\textrm{v}_T$. The VEV enters in the generation of neutrino masses via the relation $M_{ab}=\textrm{v}_T t_{ab}$, as described in \cref{sec:typeiiseesaw}. For example, $\textrm{v}_T\approx 1\GeV$ implies $m_{h^\pm}\gtrsim 130\GeV$. This mass constraint gets stronger for lower values of $\textrm{v}_T$. Therefore, the Higgs triplet accounting for neutrino masses has very stringent limits. In the ZB model \cite{Dev2017,Herrero-garcia2014}, the masses and couplings of the singly and doubly charged scalars can be independently tuned, subject to the constraint that the theory reproduce experimentally viable neutrino textures. There is therefore more flexibility in accommodating current data. In the scenario corresponding to an inverted neutrino mass hierarchy---among other assumptions---the constraints on the doubly charged scalar imply $m_{h^\pm}\geq 200\GeV$. 

\subsection{Neutrino Masses}
When considering neutrino masses, there are other sources of constraints that arise in addition to lepton flavour violation. The neutrino mass mixing matrix is related to the scalar triplet's couplings by $m_{ab}=\textrm{v}_Tt_{ab}$. Hence, the sensitivity one must achieve in $t_{ab}$ to probe the neutrino mass sector scales inversely with the VEV of the Higgs triplet. This favours using NTP to probe lower values of $\textrm{v}_T$. However, as was discussed in \cref{sec:typeiiseesaw}, this implies a larger $m_{h^\pm}$. Coupled with recent cosmological bounds on the sum of neutrino masses \cite{Couchot2017}, this makes NTP uncompetitive; we have confirmed this fact numerically. 

\section{Projected Sensitivities}\label{sec:projectedsensitivities}
Given the posterior distribution $P(\theta|\vec{x})$, we can define a 90\% C.L. interval \cite{Agashe2014}. Making use of Bayes' theorem, we can express the posterior probability in terms of a Poisson likelihood, a prior---which is a step function in the signal event rate---and a normalization. The mean of the Poisson distribution is given by $\theta'=B+S$, where $B$ is the background prediction and $S$ is the signal events. Since there is no data $\vec{x}$, we will assume that the future experiments will have observed the predicted number of background events. Collecting everything, we have
\begin{equation}
1-\alpha = \int_{-\infty}^{\theta_\text{up}} P(\theta'|\vec{x})d\theta' = 1-\frac{\Gamma(1+B,B+\theta_\text{up})}{\Gamma(1+B,B)},
\end{equation}
and solve for $\theta_\text{up}$ given $\alpha=0.1$. Setting $B+\theta_\text{up}=N_\text{SM+NP}$, we can set 90\% C.L. bounds on the couplings as a function of the masses of the new charged scalars. At SHiP, we take into account backgrounds from incoming $\nu$ and $\bar{\nu}$ whereas at DUNE, we consider only incoming $\nu$. For both collaborations, the signal dependence takes only into account incoming $\nu$. The mass of the new scalar is assumed to be in $\TeV$. \\
\begin{widetext}
\subsection{$SU(2)$ singlet scalar extensions}

\begin{table}[H]
\be
\begin{array}{ccc}
\toprule
\text{Final State} & \text{SHiP} & \text{DUNE Near Detector}\\
\midrule
e^+\mu ^-  
& 17.78\geq \frac{0.62 \left|f_{\text{$\mu $e}}\right|{}^2 \left(\left|f_{\text{e$\mu $}}\right|{}^2+\left|f_{\tau \mu
   }\right|{}^2\right)}{m_\mathcal{F}^4}-\frac{14.47 \left|f_{\text{e$\mu $}}\right|{}^2}{m_\mathcal{F}^2} 
~~~& ~~~15.53\geq \frac{0.53 \left|f_{\text{$\mu $e}}\right|{}^2 \left(\left|f_{\text{e$\mu $}}\right|{}^2+\left|f_{\tau \mu
   }\right|{}^2\right)}{m_\mathcal{F}^4}-\frac{12.66 \left|f_{\text{e$\mu $}}\right|{}^2}{m_\mathcal{F}^2} \\
e^+e^- 
& 16.82\geq -\frac{5.56 \left|f_{\text{e$\mu $}}\right|{}^2}{m_\mathcal{F}^2}+\frac{1.66 \left|f_{\text{$\mu $e}}\right|{}^2 \left(\left|f_{\text{$\mu
   $e}}\right|{}^2+\left|f_{\text{$\tau $e}}\right|{}^2\right)}{m_\mathcal{F}^4}~~~   
&~~~ 9.38\geq -\frac{4.48 \left|f_{\text{e$\mu $}}\right|{}^2}{m_\mathcal{F}^2}+\frac{1.35 \left|f_{\text{$\mu $e}}\right|{}^2 \left(\left|f_{\text{$\mu
   $e}}\right|{}^2+\left|f_{\text{$\tau $e}}\right|{}^2\right)}{m_\mathcal{F}^4}\\
\bottomrule
\end{array}
\ee
\caption{Projected 90\% C.L. sensitivity for a variety of NTP processes mediated by an $SU(2)$ singlet scalar with unit charge at both SHiP and DUNE.} 
\label{tab:constraints}
\end{table}

\subsection{$SU(2)$ triplet scalar extensions}

\begin{table}[H]
\be
\begin{array}{cc}
\toprule
\text{Final State} & \text{SHiP}\\
\midrule
e^+\mu ^-  
& 17.78\geq \frac{0.04 \left|t_{\text{ee}}\right|{}^2 \left(\left|t_{\text{e$\mu $}}\right|{}^2+\left|t_{\mu \mu }\right|{}^2+\left|t_{\tau \mu
   }\right|{}^2\right)}{m_\mathcal{T}^4}+\frac{0.16 \left|t_{\text{$\mu $e}}\right|{}^2 \left(\left|t_{\text{e$\mu $}}\right|{}^2+\left|t_{\mu \mu
   }\right|{}^2+\left|t_{\tau \mu }\right|{}^2\right)}{m_\mathcal{T}^4}+\frac{7.24 \left|t_{\text{e$\mu $}}\right|{}^2}{m_\mathcal{T}^2}\\
e^+e^- 
& 16.82\geq \frac{0.07 \left|t_{\text{ee}}\right|{}^2 \left(\left|t_{\text{ee}}\right|{}^2+\left|t_{\text{$\mu $e}}\right|{}^2+\left|t_{\text{$\tau
   $e}}\right|{}^2\right)}{m_\mathcal{T}^4}+\frac{0.42 \left|t_{\text{$\mu $e}}\right|{}^2 \left(\left|t_{\text{ee}}\right|{}^2+\left|t_{\text{$\mu
   $e}}\right|{}^2+\left|t_{\text{$\tau $e}}\right|{}^2\right)}{m_\mathcal{T}^4}+\frac{1.23 \left|t_{\text{ee}}\right|{}^2}{m_\mathcal{T}^2}-\frac{2.78
   \left|t_{\text{e$\mu $}}\right|{}^2}{m_\mathcal{T}^2}\\
\mu^+\mu^- 
& 6.43\geq \frac{0.01 \left|t_{\text{e$\mu $}}\right|{}^2 \left(\left|t_{\text{e$\mu $}}\right|{}^2+\left|t_{\mu \mu }\right|{}^2+\left|t_{\tau \mu
   }\right|{}^2\right)}{m_\mathcal{T}^4}+\frac{0.02 \left|t_{\mu \mu }\right|{}^2 \left(\left|t_{\text{e$\mu $}}\right|{}^2+\left|t_{\mu \mu
   }\right|{}^2+\left|t_{\tau \mu }\right|{}^2\right)}{m_\mathcal{T}^4}-\frac{0.04 \left|t_{\text{e$\mu $}}\right|{}^2}{m_\mathcal{T}^2}+\frac{0.28 \left|t_{\mu \mu
   }\right|{}^2}{m_\mathcal{T}^2}\\
\mu^+e^- 
& 11.65\geq \frac{0.02 \left|t_{\text{e$\mu $}}\right|{}^2 \left(\left|t_{\text{ee}}\right|{}^2+\left|t_{\text{$\mu
   $e}}\right|{}^2+\left|t_{\text{$\tau $e}}\right|{}^2\right)}{m_\mathcal{T}^4}+\frac{0.07 \left|t_{\mu \mu }\right|{}^2
   \left(\left|t_{\text{ee}}\right|{}^2+\left|t_{\text{$\mu $e}}\right|{}^2+\left|t_{\text{$\tau $e}}\right|{}^2\right)}{m_\mathcal{T}^4}-\frac{0.38
   \left|t_{\text{e$\mu $}}\right| \left|t_{\text{$\mu $e}}\right|}{m_\mathcal{T}^2}\\ 
   \bottomrule
   ~~\\
   ~~\\
\midrule
\text{Final State} & \text{DUNE Near Detector}\\
\midrule
e^+\mu ^-   
& 15.53\geq \frac{0.13 \left|t_{\text{$\mu $e}}\right|{}^2 \left(\left|t_{\text{e$\mu $}}\right|{}^2+\left|t_{\mu \mu }\right|{}^2+\left|t_{\tau \mu
   }\right|{}^2\right)}{m_\mathcal{T}^4}+\frac{6.33 \left|t_{\text{e$\mu $}}\right|{}^2}{m_\mathcal{T}^2}\\
e^+e^- 
& 9.38\geq \frac{0.34 \left|t_{\text{$\mu $e}}\right|{}^2 \left(\left|t_{\text{ee}}\right|{}^2+\left|t_{\text{$\mu
   $e}}\right|{}^2+\left|t_{\text{$\tau $e}}\right|{}^2\right)}{m_\mathcal{T}^4}+\frac{0.04 \left|t_{\text{ee}}\right|{}^2}{m_\mathcal{T}^2}-\frac{2.24
   \left|t_{\text{e$\mu $}}\right|{}^2}{m_\mathcal{T}^2}\\
\mu^+\mu^- 
& 3.9\geq \frac{0.01 \left|t_{\mu \mu }\right|{}^2 \left(\left|t_{\text{e$\mu $}}\right|{}^2+\left|t_{\mu \mu }\right|{}^2+\left|t_{\tau \mu
   }\right|{}^2\right)}{m_\mathcal{T}^4}+\frac{0.12 \left|t_{\mu \mu }\right|{}^2}{m_\mathcal{T}^2}\\
\mu^+e^- 
& 2.59\geq \frac{0.06 \left|t_{\mu \mu }\right|{}^2 \left(\left|t_{\text{ee}}\right|{}^2+\left|t_{\text{$\mu $e}}\right|{}^2+\left|t_{\text{$\tau
   $e}}\right|{}^2\right)}{m_\mathcal{T}^4}-\frac{0.01 \left|t_{\text{e$\mu $}}\right| \left|t_{\text{$\mu $e}}\right|}{m_\mathcal{T}^2}\\
   \bottomrule
\end{array}
\ee
\caption{Projected 90\% C.L. sensitivity for a variety of NTP processes mediated by the singly charged component of an $SU(2)$ triplet scalar field at both SHiP and DUNE.} 
\label{tab:constraints}
\end{table}

\end{widetext}

\bibliography{ms.bib}

\begin{thebibliography}{56}%
\makeatletter
\providecommand \@ifxundefined [1]{%
 \@ifx{#1\undefined}
}%
\providecommand \@ifnum [1]{%
 \ifnum #1\expandafter \@firstoftwo
 \else \expandafter \@secondoftwo
 \fi
}%
\providecommand \@ifx [1]{%
 \ifx #1\expandafter \@firstoftwo
 \else \expandafter \@secondoftwo
 \fi
}%
\providecommand \natexlab [1]{#1}%
\providecommand \enquote  [1]{``#1''}%
\providecommand \bibnamefont  [1]{#1}%
\providecommand \bibfnamefont [1]{#1}%
\providecommand \citenamefont [1]{#1}%
\providecommand \href@noop [0]{\@secondoftwo}%
\providecommand \href [0]{\begingroup \@sanitize@url \@href}%
\providecommand \@href[1]{\@@startlink{#1}\@@href}%
\providecommand \@@href[1]{\endgroup#1\@@endlink}%
\providecommand \@sanitize@url [0]{\catcode `\\12\catcode `\$12\catcode
  `\&12\catcode `\#12\catcode `\^12\catcode `\_12\catcode `\%12\relax}%
\providecommand \@@startlink[1]{}%
\providecommand \@@endlink[0]{}%
\providecommand \url  [0]{\begingroup\@sanitize@url \@url }%
\providecommand \@url [1]{\endgroup\@href {#1}{\urlprefix }}%
\providecommand \urlprefix  [0]{URL }%
\providecommand \Eprint [0]{\href }%
\providecommand \doibase [0]{http://dx.doi.org/}%
\providecommand \selectlanguage [0]{\@gobble}%
\providecommand \bibinfo  [0]{\@secondoftwo}%
\providecommand \bibfield  [0]{\@secondoftwo}%
\providecommand \translation [1]{[#1]}%
\providecommand \BibitemOpen [0]{}%
\providecommand \bibitemStop [0]{}%
\providecommand \bibitemNoStop [0]{.\EOS\space}%
\providecommand \EOS [0]{\spacefactor3000\relax}%
\providecommand \BibitemShut  [1]{\csname bibitem#1\endcsname}%
\let\auto@bib@innerbib\@empty
\bibitem [{\citenamefont {Pohl}\ \emph {et~al.}(2013)\citenamefont {Pohl},
  \citenamefont {Gilman}, \citenamefont {Miller},\ and\ \citenamefont
  {Pachucki}}]{Pohl2013}%
  \BibitemOpen
  \bibfield  {author} {\bibinfo {author} {\bibfnamefont {Randolf}\ \bibnamefont
  {Pohl}}, \bibinfo {author} {\bibfnamefont {Ronald}\ \bibnamefont {Gilman}},
  \bibinfo {author} {\bibfnamefont {Gerald~A.}\ \bibnamefont {Miller}}, \ and\
  \bibinfo {author} {\bibfnamefont {Krzysztof}\ \bibnamefont {Pachucki}},\
  }\bibfield  {title} {\enquote {\bibinfo {title} {{Muonic hydrogen and the
  proton radius puzzle}},}\ }\href {\doibase
  10.1146/annurev-nucl-102212-170627} {\bibfield  {journal} {\bibinfo
  {journal} {Ann. Rev. Nucl. Part. Sci.}\ }\textbf {\bibinfo {volume} {63}},\
  \bibinfo {pages} {175--204} (\bibinfo {year} {2013})},\ \Eprint
  {http://arxiv.org/abs/1301.0905} {arXiv:1301.0905 [physics.atom-ph]}
  \BibitemShut {NoStop}%
\bibitem [{\citenamefont {Hill}(2017)}]{Hill2017}%
  \BibitemOpen
  \bibfield  {author} {\bibinfo {author} {\bibfnamefont {Richard~J.}\
  \bibnamefont {Hill}},\ }\bibfield  {title} {\enquote {\bibinfo {title}
  {{Review of experimental and theoretical status of the proton radius
  puzzle}},}\ }\bibfield  {booktitle} {\emph {\bibinfo {booktitle}
  {{Proceedings, 12th Conference on Quark Confinement and the Hadron Spectrum
  (Confinement XII): Thessaloniki, Greece}}},\ }\href {\doibase
  10.1051/epjconf/201713701023} {\bibfield  {journal} {\bibinfo  {journal} {EPJ
  Web Conf.}\ }\textbf {\bibinfo {volume} {137}},\ \bibinfo {pages} {01023}
  (\bibinfo {year} {2017})},\ \Eprint {http://arxiv.org/abs/1702.01189}
  {arXiv:1702.01189 [hep-ph]} \BibitemShut {NoStop}%
\bibitem [{\citenamefont {Jegerlehner}\ and\ \citenamefont
  {Nyffeler}(2009)}]{Jegerlehner:2009ry}%
  \BibitemOpen
  \bibfield  {author} {\bibinfo {author} {\bibfnamefont {Fred}\ \bibnamefont
  {Jegerlehner}}\ and\ \bibinfo {author} {\bibfnamefont {Andreas}\ \bibnamefont
  {Nyffeler}},\ }\bibfield  {title} {\enquote {\bibinfo {title} {{The Muon
  g-2}},}\ }\href {\doibase 10.1016/j.physrep.2009.04.003} {\bibfield
  {journal} {\bibinfo  {journal} {Phys. Rept.}\ }\textbf {\bibinfo {volume}
  {477}},\ \bibinfo {pages} {1--110} (\bibinfo {year} {2009})},\ \Eprint
  {http://arxiv.org/abs/0902.3360} {arXiv:0902.3360 [hep-ph]} \BibitemShut
  {NoStop}%
\bibitem [{\citenamefont {Bennett}\ \emph {et~al.}(2006)\citenamefont {Bennett}
  \emph {et~al.}}]{Bennett2006}%
  \BibitemOpen
  \bibfield  {author} {\bibinfo {author} {\bibfnamefont {G.~W.}\ \bibnamefont
  {Bennett}} \emph {et~al.} (\bibinfo {collaboration} {Muon g-2}),\ }\bibfield
  {title} {\enquote {\bibinfo {title} {{Final Report of the Muon E821 Anomalous
  Magnetic Moment Measurement at BNL}},}\ }\href {\doibase
  10.1103/PhysRevD.73.072003} {\bibfield  {journal} {\bibinfo  {journal} {Phys.
  Rev.}\ }\textbf {\bibinfo {volume} {D73}},\ \bibinfo {pages} {072003}
  (\bibinfo {year} {2006})},\ \Eprint {http://arxiv.org/abs/hep-ex/0602035}
  {arXiv:hep-ex/0602035 [hep-ex]} \BibitemShut {NoStop}%
\bibitem [{\citenamefont {Aguilar-Arevalo}\ \emph {et~al.}(2001)\citenamefont
  {Aguilar-Arevalo} \emph {et~al.}}]{Aguilar2001}%
  \BibitemOpen
  \bibfield  {author} {\bibinfo {author} {\bibfnamefont {A.}~\bibnamefont
  {Aguilar-Arevalo}} \emph {et~al.} (\bibinfo {collaboration} {LSND}),\
  }\bibfield  {title} {\enquote {\bibinfo {title} {{Evidence for neutrino
  oscillations from the observation of anti-neutrino(electron) appearance in a
  anti-neutrino(muon) beam}},}\ }\href {\doibase 10.1103/PhysRevD.64.112007}
  {\bibfield  {journal} {\bibinfo  {journal} {Phys. Rev.}\ }\textbf {\bibinfo
  {volume} {D64}},\ \bibinfo {pages} {112007} (\bibinfo {year} {2001})},\
  \Eprint {http://arxiv.org/abs/hep-ex/0104049} {arXiv:hep-ex/0104049 [hep-ex]}
  \BibitemShut {NoStop}%
\bibitem [{\citenamefont {Babu}(1988)}]{Babu1988}%
  \BibitemOpen
  \bibfield  {author} {\bibinfo {author} {\bibfnamefont {K.~S.}\ \bibnamefont
  {Babu}},\ }\bibfield  {title} {\enquote {\bibinfo {title} {{Model of
  `Calculable' Majorana Neutrino Masses}},}\ }\href {\doibase
  10.1016/0370-2693(88)91584-5} {\bibfield  {journal} {\bibinfo  {journal}
  {Phys. Lett.}\ }\textbf {\bibinfo {volume} {B203}},\ \bibinfo {pages}
  {132--136} (\bibinfo {year} {1988})}\BibitemShut {NoStop}%
\bibitem [{\citenamefont {Babu}\ and\ \citenamefont
  {Pakvasa}(2002)}]{Babu2002}%
  \BibitemOpen
  \bibfield  {author} {\bibinfo {author} {\bibfnamefont {K.~S.}\ \bibnamefont
  {Babu}}\ and\ \bibinfo {author} {\bibfnamefont {Sandip}\ \bibnamefont
  {Pakvasa}},\ }\bibfield  {title} {\enquote {\bibinfo {title} {{Lepton number
  violating muon decay and the LSND neutrino anomaly}},}\ }\href@noop {} {\
  (\bibinfo {year} {2002})},\ \Eprint {http://arxiv.org/abs/hep-ph/0204236}
  {arXiv:hep-ph/0204236 [hep-ph]} \BibitemShut {NoStop}%
\bibitem [{\citenamefont {Lindner}\ \emph {et~al.}(2016)\citenamefont
  {Lindner}, \citenamefont {Platscher},\ and\ \citenamefont
  {Queiroz}}]{Lindner2016}%
  \BibitemOpen
  \bibfield  {author} {\bibinfo {author} {\bibfnamefont {Manfred}\ \bibnamefont
  {Lindner}}, \bibinfo {author} {\bibfnamefont {Moritz}\ \bibnamefont
  {Platscher}}, \ and\ \bibinfo {author} {\bibfnamefont {Farinaldo~S.}\
  \bibnamefont {Queiroz}},\ }\bibfield  {title} {\enquote {\bibinfo {title} {{A
  Call for New Physics : The Muon Anomalous Magnetic Moment and Lepton Flavor
  Violation}},}\ }\href@noop {} {\  (\bibinfo {year} {2016})},\ \Eprint
  {http://arxiv.org/abs/1610.06587} {arXiv:1610.06587 [hep-ph]} \BibitemShut
  {NoStop}%
\bibitem [{\citenamefont {Liu}\ \emph {et~al.}(2016)\citenamefont {Liu},
  \citenamefont {McKeen},\ and\ \citenamefont {Miller}}]{Liu2016}%
  \BibitemOpen
  \bibfield  {author} {\bibinfo {author} {\bibfnamefont {Yu-Sheng}\
  \bibnamefont {Liu}}, \bibinfo {author} {\bibfnamefont {David}\ \bibnamefont
  {McKeen}}, \ and\ \bibinfo {author} {\bibfnamefont {Gerald~A.}\ \bibnamefont
  {Miller}},\ }\bibfield  {title} {\enquote {\bibinfo {title} {{Electrophobic
  Scalar Boson and Muonic Puzzles}},}\ }\href {\doibase
  10.1103/PhysRevLett.117.101801} {\bibfield  {journal} {\bibinfo  {journal}
  {Phys. Rev. Lett.}\ }\textbf {\bibinfo {volume} {117}},\ \bibinfo {pages}
  {101801} (\bibinfo {year} {2016})},\ \Eprint
  {http://arxiv.org/abs/1605.04612} {arXiv:1605.04612 [hep-ph]} \BibitemShut
  {NoStop}%
\bibitem [{\citenamefont {Herrero-Garcia}\ \emph {et~al.}(2014)\citenamefont
  {Herrero-Garcia}, \citenamefont {Nebot}, \citenamefont {Rius},\ and\
  \citenamefont {Santamaria}}]{Herrero-garcia2014}%
  \BibitemOpen
  \bibfield  {author} {\bibinfo {author} {\bibfnamefont {Juan}\ \bibnamefont
  {Herrero-Garcia}}, \bibinfo {author} {\bibfnamefont {Miguel}\ \bibnamefont
  {Nebot}}, \bibinfo {author} {\bibfnamefont {Nuria}\ \bibnamefont {Rius}}, \
  and\ \bibinfo {author} {\bibfnamefont {Arcadi}\ \bibnamefont {Santamaria}},\
  }\bibfield  {title} {\enquote {\bibinfo {title} {{The Zee-Babu model
  revisited in the light of new data}},}\ }\href {\doibase
  10.1016/j.nuclphysb.2014.06.001} {\bibfield  {journal} {\bibinfo  {journal}
  {Nucl. Phys.}\ }\textbf {\bibinfo {volume} {B885}},\ \bibinfo {pages}
  {542--570} (\bibinfo {year} {2014})},\ \Eprint
  {http://arxiv.org/abs/1402.4491} {arXiv:1402.4491 [hep-ph]} \BibitemShut
  {NoStop}%
\bibitem [{\citenamefont {Dev}\ \emph {et~al.}(2017)\citenamefont {Dev},
  \citenamefont {Vila},\ and\ \citenamefont {Rodejohann}}]{Dev2017}%
  \BibitemOpen
  \bibfield  {author} {\bibinfo {author} {\bibfnamefont {P.~S.~Bhupal}\
  \bibnamefont {Dev}}, \bibinfo {author} {\bibfnamefont {Clara~Miralles}\
  \bibnamefont {Vila}}, \ and\ \bibinfo {author} {\bibfnamefont {Werner}\
  \bibnamefont {Rodejohann}},\ }\href@noop {} {\enquote {\bibinfo {title}
  {{Naturalness in testable type II seesaw scenarios}},}\ } (\bibinfo {year}
  {2017}),\ \Eprint {http://arxiv.org/abs/1703.00828} {arXiv:1703.00828
  [hep-ph]} \BibitemShut {NoStop}%
\bibitem [{\citenamefont {Altmannshofer}\ \emph {et~al.}(2014)\citenamefont
  {Altmannshofer}, \citenamefont {Gori}, \citenamefont {Pospelov},\ and\
  \citenamefont {Yavin}}]{Altmannshofer2014}%
  \BibitemOpen
  \bibfield  {author} {\bibinfo {author} {\bibfnamefont {Wolfgang}\
  \bibnamefont {Altmannshofer}}, \bibinfo {author} {\bibfnamefont {Stefania}\
  \bibnamefont {Gori}}, \bibinfo {author} {\bibfnamefont {Maxim}\ \bibnamefont
  {Pospelov}}, \ and\ \bibinfo {author} {\bibfnamefont {Itay}\ \bibnamefont
  {Yavin}},\ }\bibfield  {title} {\enquote {\bibinfo {title} {{Neutrino Trident
  Production: A Powerful Probe of New Physics with Neutrino Beams}},}\ }\href
  {\doibase 10.1103/PhysRevLett.113.091801} {\bibfield  {journal} {\bibinfo
  {journal} {Phys. Rev. Lett.}\ }\textbf {\bibinfo {volume} {113}},\ \bibinfo
  {pages} {091801} (\bibinfo {year} {2014})},\ \Eprint
  {http://arxiv.org/abs/1406.2332} {arXiv:1406.2332 [hep-ph]} \BibitemShut
  {NoStop}%
\bibitem [{\citenamefont {Geiregat}\ \emph {et~al.}(1990)\citenamefont
  {Geiregat} \emph {et~al.}}]{Geiregat1990}%
  \BibitemOpen
  \bibfield  {author} {\bibinfo {author} {\bibfnamefont {D.}~\bibnamefont
  {Geiregat}} \emph {et~al.} (\bibinfo {collaboration} {CHARM-II}),\ }\bibfield
   {title} {\enquote {\bibinfo {title} {{First observation of neutrino trident
  production}},}\ }\href {\doibase 10.1016/0370-2693(90)90146-W} {\bibfield
  {journal} {\bibinfo  {journal} {Phys. Lett.}\ }\textbf {\bibinfo {volume}
  {B245}},\ \bibinfo {pages} {271--275} (\bibinfo {year} {1990})}\BibitemShut
  {NoStop}%
\bibitem [{\citenamefont {Mishra}\ \emph {et~al.}(1991)\citenamefont {Mishra}
  \emph {et~al.}}]{Mishra1991}%
  \BibitemOpen
  \bibfield  {author} {\bibinfo {author} {\bibfnamefont {S.~R.}\ \bibnamefont
  {Mishra}} \emph {et~al.} (\bibinfo {collaboration} {CCFR}),\ }\bibfield
  {title} {\enquote {\bibinfo {title} {{Neutrino tridents and W Z
  interference}},}\ }\href {\doibase 10.1103/PhysRevLett.66.3117} {\bibfield
  {journal} {\bibinfo  {journal} {Phys. Rev. Lett.}\ }\textbf {\bibinfo
  {volume} {66}},\ \bibinfo {pages} {3117--3120} (\bibinfo {year}
  {1991})}\BibitemShut {NoStop}%
\bibitem [{\citenamefont {Magill}\ and\ \citenamefont
  {Plestid}(2017)}]{Magill2016}%
  \BibitemOpen
  \bibfield  {author} {\bibinfo {author} {\bibfnamefont {Gabriel}\ \bibnamefont
  {Magill}}\ and\ \bibinfo {author} {\bibfnamefont {Ryan}\ \bibnamefont
  {Plestid}},\ }\bibfield  {title} {\enquote {\bibinfo {title} {{Neutrino
  Trident Production at the Intensity Frontier}},}\ }\href {\doibase
  10.1103/PhysRevD.95.073004} {\bibfield  {journal} {\bibinfo  {journal} {Phys.
  Rev.}\ }\textbf {\bibinfo {volume} {D95}},\ \bibinfo {pages} {073004}
  (\bibinfo {year} {2017})},\ \Eprint {http://arxiv.org/abs/1612.05642}
  {arXiv:1612.05642 [hep-ph]} \BibitemShut {NoStop}%
\bibitem [{\citenamefont {Acciarri}\ \emph {et~al.}(2015)\citenamefont
  {Acciarri} \emph {et~al.}}]{DUNECollaboration2015}%
  \BibitemOpen
  \bibfield  {author} {\bibinfo {author} {\bibfnamefont {R.}~\bibnamefont
  {Acciarri}} \emph {et~al.} (\bibinfo {collaboration} {DUNE}),\ }\bibfield
  {title} {\enquote {\bibinfo {title} {{Long-Baseline Neutrino Facility (LBNF)
  and Deep Underground Neutrino Experiment (DUNE)}},}\ }\href@noop {} {\
  (\bibinfo {year} {2015})},\ \Eprint {http://arxiv.org/abs/1512.06148}
  {arXiv:1512.06148 [physics.ins-det]} \BibitemShut {NoStop}%
\bibitem [{\citenamefont {Anelli}\ \emph {et~al.}(2015)\citenamefont {Anelli}
  \emph {et~al.}}]{SHiPCollaboration2015}%
  \BibitemOpen
  \bibfield  {author} {\bibinfo {author} {\bibfnamefont {M.}~\bibnamefont
  {Anelli}} \emph {et~al.} (\bibinfo {collaboration} {SHiP}),\ }\bibfield
  {title} {\enquote {\bibinfo {title} {{A facility to Search for Hidden
  Particles (SHiP) at the CERN SPS}},}\ }\href@noop {} {\  (\bibinfo {year}
  {2015})},\ \Eprint {http://arxiv.org/abs/1504.04956} {arXiv:1504.04956
  [physics.ins-det]} \BibitemShut {NoStop}%
\bibitem [{\citenamefont {Chen}\ \emph {et~al.}(2007)\citenamefont {Chen} \emph
  {et~al.}}]{Chen2007}%
  \BibitemOpen
  \bibfield  {author} {\bibinfo {author} {\bibfnamefont {H.}~\bibnamefont
  {Chen}} \emph {et~al.} (\bibinfo {collaboration} {MicroBooNE}),\ }\bibfield
  {title} {\enquote {\bibinfo {title} {{Proposal for a New Experiment Using the
  Booster and NuMI Neutrino Beamlines: MicroBooNE}},}\ }\href@noop {} {\
  (\bibinfo {year} {2007})}\BibitemShut {NoStop}%
\bibitem [{\citenamefont {Belusevic}\ and\ \citenamefont
  {Smith}(1988)}]{Belusevic1988}%
  \BibitemOpen
  \bibfield  {author} {\bibinfo {author} {\bibfnamefont {R.}~\bibnamefont
  {Belusevic}}\ and\ \bibinfo {author} {\bibfnamefont {J.}~\bibnamefont
  {Smith}},\ }\bibfield  {title} {\enquote {\bibinfo {title} {W-z interference
  in \ensuremath{\nu}-nucleus scattering},}\ }\href {\doibase
  10.1103/PhysRevD.37.2419} {\bibfield  {journal} {\bibinfo  {journal} {Phys.
  Rev. D}\ }\textbf {\bibinfo {volume} {37}},\ \bibinfo {pages} {2419--2422}
  (\bibinfo {year} {1988})}\BibitemShut {NoStop}%
\bibitem [{\citenamefont {Lovseth}\ and\ \citenamefont
  {Radomiski}(1971)}]{Lazvseth1971}%
  \BibitemOpen
  \bibfield  {author} {\bibinfo {author} {\bibfnamefont {J.}~\bibnamefont
  {Lovseth}}\ and\ \bibinfo {author} {\bibfnamefont {M.}~\bibnamefont
  {Radomiski}},\ }\bibfield  {title} {\enquote {\bibinfo {title} {{Kinematical
  distributions of neutrino-produced lepton triplets}},}\ }\href {\doibase
  10.1103/PhysRevD.3.2686} {\bibfield  {journal} {\bibinfo  {journal} {Phys.
  Rev.}\ }\textbf {\bibinfo {volume} {D3}},\ \bibinfo {pages} {2686--2706}
  (\bibinfo {year} {1971})}\BibitemShut {NoStop}%
\bibitem [{\citenamefont {Brown}\ \emph {et~al.}(1972)\citenamefont {Brown},
  \citenamefont {Hobbs}, \citenamefont {Smith},\ and\ \citenamefont
  {Stanko}}]{Brown1972}%
  \BibitemOpen
  \bibfield  {author} {\bibinfo {author} {\bibfnamefont {R.~W.}\ \bibnamefont
  {Brown}}, \bibinfo {author} {\bibfnamefont {R.~H.}\ \bibnamefont {Hobbs}},
  \bibinfo {author} {\bibfnamefont {J.}~\bibnamefont {Smith}}, \ and\ \bibinfo
  {author} {\bibfnamefont {N.}~\bibnamefont {Stanko}},\ }\bibfield  {title}
  {\enquote {\bibinfo {title} {{Intermediate boson. iii. virtual-boson effects
  in neutrino trident production}},}\ }\href {\doibase 10.1103/PhysRevD.6.3273}
  {\bibfield  {journal} {\bibinfo  {journal} {Phys. Rev.}\ }\textbf {\bibinfo
  {volume} {D6}},\ \bibinfo {pages} {3273--3292} (\bibinfo {year}
  {1972})}\BibitemShut {NoStop}%
\bibitem [{\citenamefont {Ge}\ \emph {et~al.}(2017)\citenamefont {Ge},
  \citenamefont {Lindner},\ and\ \citenamefont {Rodejohann}}]{Ge:2017poy}%
  \BibitemOpen
  \bibfield  {author} {\bibinfo {author} {\bibfnamefont {Shao-Feng}\
  \bibnamefont {Ge}}, \bibinfo {author} {\bibfnamefont {Manfred}\ \bibnamefont
  {Lindner}}, \ and\ \bibinfo {author} {\bibfnamefont {Werner}\ \bibnamefont
  {Rodejohann}},\ }\bibfield  {title} {\enquote {\bibinfo {title} {{Atmospheric
  Trident Production for Probing New Physics}},}\ }\href {\doibase
  10.1016/j.physletb.2017.06.020} {\bibfield  {journal} {\bibinfo  {journal}
  {Phys. Lett.}\ }\textbf {\bibinfo {volume} {B772}},\ \bibinfo {pages}
  {164--168} (\bibinfo {year} {2017})},\ \Eprint
  {http://arxiv.org/abs/1702.02617} {arXiv:1702.02617 [hep-ph]} \BibitemShut
  {NoStop}%
\bibitem [{\citenamefont {Dreiner}\ \emph {et~al.}(2010)\citenamefont
  {Dreiner}, \citenamefont {Haber},\ and\ \citenamefont
  {Martin}}]{Dreiner:2008tw}%
  \BibitemOpen
  \bibfield  {author} {\bibinfo {author} {\bibfnamefont {Herbi~K.}\
  \bibnamefont {Dreiner}}, \bibinfo {author} {\bibfnamefont {Howard~E.}\
  \bibnamefont {Haber}}, \ and\ \bibinfo {author} {\bibfnamefont {Stephen~P.}\
  \bibnamefont {Martin}},\ }\bibfield  {title} {\enquote {\bibinfo {title}
  {{Two-component spinor techniques and Feynman rules for quantum field theory
  and supersymmetry}},}\ }\href {\doibase 10.1016/j.physrep.2010.05.002}
  {\bibfield  {journal} {\bibinfo  {journal} {Phys. Rept.}\ }\textbf {\bibinfo
  {volume} {494}},\ \bibinfo {pages} {1--196} (\bibinfo {year} {2010})},\
  \Eprint {http://arxiv.org/abs/0812.1594} {arXiv:0812.1594 [hep-ph]}
  \BibitemShut {NoStop}%
\bibitem [{\citenamefont {Budnev}\ \emph {et~al.}(1975)\citenamefont {Budnev},
  \citenamefont {Ginzburg}, \citenamefont {Meledin},\ and\ \citenamefont
  {Serbo}}]{Budnev1975}%
  \BibitemOpen
  \bibfield  {author} {\bibinfo {author} {\bibfnamefont {V.~M.}\ \bibnamefont
  {Budnev}}, \bibinfo {author} {\bibfnamefont {I.~F.}\ \bibnamefont
  {Ginzburg}}, \bibinfo {author} {\bibfnamefont {G.~V.}\ \bibnamefont
  {Meledin}}, \ and\ \bibinfo {author} {\bibfnamefont {V.~G.}\ \bibnamefont
  {Serbo}},\ }\bibfield  {title} {\enquote {\bibinfo {title} {{The Two photon
  particle production mechanism. Physical problems. Applications. Equivalent
  photon approximation}},}\ }\href {\doibase 10.1016/0370-1573(75)90009-5}
  {\bibfield  {journal} {\bibinfo  {journal} {Phys. Rept.}\ }\textbf {\bibinfo
  {volume} {15}},\ \bibinfo {pages} {181--281} (\bibinfo {year}
  {1975})}\BibitemShut {NoStop}%
\bibitem [{\citenamefont {Vermaseren}(2000)}]{Manipulation}%
  \BibitemOpen
  \bibfield  {author} {\bibinfo {author} {\bibfnamefont {J.~A.~M.}\
  \bibnamefont {Vermaseren}},\ }\bibfield  {title} {\enquote {\bibinfo {title}
  {{New features of FORM}},}\ }\href@noop {} {\  (\bibinfo {year} {2000})},\
  \Eprint {http://arxiv.org/abs/math-ph/0010025} {arXiv:math-ph/0010025
  [math-ph]} \BibitemShut {NoStop}%
\bibitem [{\citenamefont {Abbiendi}\ \emph {et~al.}(2013)\citenamefont
  {Abbiendi} \emph {et~al.}}]{Aleph2013}%
  \BibitemOpen
  \bibfield  {author} {\bibinfo {author} {\bibfnamefont {G.}~\bibnamefont
  {Abbiendi}} \emph {et~al.} (\bibinfo {collaboration} {LEP, DELPHI, OPAL,
  ALEPH, L3}),\ }\bibfield  {title} {\enquote {\bibinfo {title} {{Search for
  Charged Higgs bosons: Combined Results Using LEP Data}},}\ }\href {\doibase
  10.1140/epjc/s10052-013-2463-1} {\bibfield  {journal} {\bibinfo  {journal}
  {Eur. Phys. J.}\ }\textbf {\bibinfo {volume} {C73}},\ \bibinfo {pages} {2463}
  (\bibinfo {year} {2013})},\ \Eprint {http://arxiv.org/abs/1301.6065}
  {arXiv:1301.6065 [hep-ex]} \BibitemShut {NoStop}%
\bibitem [{\citenamefont {Jentschura}\ and\ \citenamefont
  {Serbo}(2009)}]{Jentschura2009}%
  \BibitemOpen
  \bibfield  {author} {\bibinfo {author} {\bibfnamefont {U.~D.}\ \bibnamefont
  {Jentschura}}\ and\ \bibinfo {author} {\bibfnamefont {V.~G.}\ \bibnamefont
  {Serbo}},\ }\bibfield  {title} {\enquote {\bibinfo {title} {{Nuclear form
  factor, validity of the equivalent photon approximation and Coulomb
  corrections to muon pair production in photon-nucleus and nucleus-nucleus
  collisions}},}\ }\href {\doibase 10.1140/epjc/s10052-009-1147-3} {\bibfield
  {journal} {\bibinfo  {journal} {Eur. Phys. J.}\ }\textbf {\bibinfo {volume}
  {C64}},\ \bibinfo {pages} {309--317} (\bibinfo {year} {2009})},\ \Eprint
  {http://arxiv.org/abs/0908.3853} {arXiv:0908.3853 [hep-ph]} \BibitemShut
  {NoStop}%
\bibitem [{\citenamefont {Nebot}\ \emph {et~al.}(2008)\citenamefont {Nebot},
  \citenamefont {Oliver}, \citenamefont {Palao},\ and\ \citenamefont
  {Santamaria}}]{Nebot2007}%
  \BibitemOpen
  \bibfield  {author} {\bibinfo {author} {\bibfnamefont {Miguel}\ \bibnamefont
  {Nebot}}, \bibinfo {author} {\bibfnamefont {Josep~F.}\ \bibnamefont
  {Oliver}}, \bibinfo {author} {\bibfnamefont {David}\ \bibnamefont {Palao}}, \
  and\ \bibinfo {author} {\bibfnamefont {Arcadi}\ \bibnamefont {Santamaria}},\
  }\bibfield  {title} {\enquote {\bibinfo {title} {{Prospects for the Zee-Babu
  Model at the CERN LHC and low energy experiments}},}\ }\href {\doibase
  10.1103/PhysRevD.77.093013} {\bibfield  {journal} {\bibinfo  {journal} {Phys.
  Rev.}\ }\textbf {\bibinfo {volume} {D77}},\ \bibinfo {pages} {093013}
  (\bibinfo {year} {2008})},\ \Eprint {http://arxiv.org/abs/0711.0483}
  {arXiv:0711.0483 [hep-ph]} \BibitemShut {NoStop}%
\bibitem [{\citenamefont {Agafonova}\ \emph {et~al.}(2015)\citenamefont
  {Agafonova} \emph {et~al.}}]{Agafonova2015}%
  \BibitemOpen
  \bibfield  {author} {\bibinfo {author} {\bibfnamefont {N.}~\bibnamefont
  {Agafonova}} \emph {et~al.} (\bibinfo {collaboration} {OPERA}),\ }\bibfield
  {title} {\enquote {\bibinfo {title} {{Discovery of $\tau$ Neutrino Appearance
  in the CNGS Neutrino Beam with the OPERA Experiment}},}\ }\href {\doibase
  10.1103/PhysRevLett.115.121802} {\bibfield  {journal} {\bibinfo  {journal}
  {Phys. Rev. Lett.}\ }\textbf {\bibinfo {volume} {115}},\ \bibinfo {pages}
  {121802} (\bibinfo {year} {2015})},\ \Eprint
  {http://arxiv.org/abs/1507.01417} {arXiv:1507.01417 [hep-ex]} \BibitemShut
  {NoStop}%
\bibitem [{\citenamefont {King}\ \emph {et~al.}(1991)\citenamefont {King} \emph
  {et~al.}}]{KING1991254}%
  \BibitemOpen
  \bibfield  {author} {\bibinfo {author} {\bibfnamefont {B.~J.}\ \bibnamefont
  {King}} \emph {et~al.},\ }\bibfield  {title} {\enquote {\bibinfo {title}
  {{Measuring muon momenta with the CCFR neutrino detector}},}\ }\href
  {\doibase 10.1016/0168-9002(91)90408-I} {\bibfield  {journal} {\bibinfo
  {journal} {Nucl. Instrum. Meth.}\ }\textbf {\bibinfo {volume} {A302}},\
  \bibinfo {pages} {254--260} (\bibinfo {year} {1991})}\BibitemShut {NoStop}%
\bibitem [{\citenamefont {Adams}\ \emph {et~al.}(2000)\citenamefont {Adams}
  \emph {et~al.}}]{Adams:1999mn}%
  \BibitemOpen
  \bibfield  {author} {\bibinfo {author} {\bibfnamefont {T.}~\bibnamefont
  {Adams}} \emph {et~al.} (\bibinfo {collaboration} {NuTeV}),\ }\bibfield
  {title} {\enquote {\bibinfo {title} {{Evidence for diffractive charm
  production in muon-neutrino Fe and anti-muon-neutrino Fe scattering at the
  Tevatron}},}\ }\href {\doibase 10.1103/PhysRevD.61.092001} {\bibfield
  {journal} {\bibinfo  {journal} {Phys. Rev.}\ }\textbf {\bibinfo {volume}
  {D61}},\ \bibinfo {pages} {092001} (\bibinfo {year} {2000})},\ \Eprint
  {http://arxiv.org/abs/hep-ex/9909041} {arXiv:hep-ex/9909041 [hep-ex]}
  \BibitemShut {NoStop}%
\bibitem [{\citenamefont {Buonaura}(2016)}]{Buonaura2015}%
  \BibitemOpen
  \bibfield  {author} {\bibinfo {author} {\bibfnamefont {Annarita}\
  \bibnamefont {Buonaura}} (\bibinfo {collaboration} {SHiP Collaboration}),\
  }\bibfield  {title} {\enquote {\bibinfo {title} {{The SHiP experiment and its
  detector for neutrino physics}},}\ }\href
  {https://cds.cern.ch/record/2129493} {\  (\bibinfo {year}
  {2016})}\BibitemShut {NoStop}%
\bibitem [{\citenamefont {Cao}\ \emph {et~al.}(2009)\citenamefont {Cao},
  \citenamefont {Wan}, \citenamefont {Wu},\ and\ \citenamefont
  {Yang}}]{Cao2009}%
  \BibitemOpen
  \bibfield  {author} {\bibinfo {author} {\bibfnamefont {Junjie}\ \bibnamefont
  {Cao}}, \bibinfo {author} {\bibfnamefont {Peihua}\ \bibnamefont {Wan}},
  \bibinfo {author} {\bibfnamefont {Lei}\ \bibnamefont {Wu}}, \ and\ \bibinfo
  {author} {\bibfnamefont {Jin~Min}\ \bibnamefont {Yang}},\ }\bibfield  {title}
  {\enquote {\bibinfo {title} {{Lepton-Specific Two-Higgs Doublet Model:
  Experimental Constraints and Implication on Higgs Phenomenology}},}\ }\href
  {\doibase 10.1103/PhysRevD.80.071701} {\bibfield  {journal} {\bibinfo
  {journal} {Phys. Rev.}\ }\textbf {\bibinfo {volume} {D80}},\ \bibinfo {pages}
  {071701} (\bibinfo {year} {2009})},\ \Eprint {http://arxiv.org/abs/0909.5148}
  {arXiv:0909.5148 [hep-ph]} \BibitemShut {NoStop}%
\bibitem [{\citenamefont {Esteban}\ \emph {et~al.}(2017)\citenamefont
  {Esteban}, \citenamefont {Gonzalez-Garcia}, \citenamefont {Maltoni},
  \citenamefont {Martinez-Soler},\ and\ \citenamefont {Schwetz}}]{Esteban2017}%
  \BibitemOpen
  \bibfield  {author} {\bibinfo {author} {\bibfnamefont {Ivan}\ \bibnamefont
  {Esteban}}, \bibinfo {author} {\bibfnamefont {M.~C.}\ \bibnamefont
  {Gonzalez-Garcia}}, \bibinfo {author} {\bibfnamefont {Michele}\ \bibnamefont
  {Maltoni}}, \bibinfo {author} {\bibfnamefont {Ivan}\ \bibnamefont
  {Martinez-Soler}}, \ and\ \bibinfo {author} {\bibfnamefont {Thomas}\
  \bibnamefont {Schwetz}},\ }\bibfield  {title} {\enquote {\bibinfo {title}
  {Updated fit to three neutrino mixing: exploring the accelerator-reactor
  complementarity},}\ }\href {\doibase 10.1007/JHEP01(2017)087} {\bibfield
  {journal} {\bibinfo  {journal} {Journal of High Energy Physics}\ }\textbf
  {\bibinfo {volume} {2017}},\ \bibinfo {pages} {87} (\bibinfo {year}
  {2017})}\BibitemShut {NoStop}%
\bibitem [{\citenamefont {Abe}\ \emph {et~al.}(2015)\citenamefont {Abe},
  \citenamefont {Sato},\ and\ \citenamefont {Yagyu}}]{Abe2015}%
  \BibitemOpen
  \bibfield  {author} {\bibinfo {author} {\bibfnamefont {Tomohiro}\
  \bibnamefont {Abe}}, \bibinfo {author} {\bibfnamefont {Ryosuke}\ \bibnamefont
  {Sato}}, \ and\ \bibinfo {author} {\bibfnamefont {Kei}\ \bibnamefont
  {Yagyu}},\ }\bibfield  {title} {\enquote {\bibinfo {title} {{Lepton-specific
  two Higgs doublet model as a solution of muon g-2 anomaly}},}\ }\href
  {\doibase 10.1007/JHEP07(2015)064} {\bibfield  {journal} {\bibinfo  {journal}
  {JHEP}\ }\textbf {\bibinfo {volume} {07}},\ \bibinfo {pages} {064} (\bibinfo
  {year} {2015})},\ \Eprint {http://arxiv.org/abs/1504.07059} {arXiv:1504.07059
  [hep-ph]} \BibitemShut {NoStop}%
\bibitem [{\citenamefont {Branco}\ \emph {et~al.}(2012)\citenamefont {Branco},
  \citenamefont {Ferreira}, \citenamefont {Lavoura}, \citenamefont {Rebelo},
  \citenamefont {Sher},\ and\ \citenamefont {Silva}}]{Branco:2011iw}%
  \BibitemOpen
  \bibfield  {author} {\bibinfo {author} {\bibfnamefont {G.~C.}\ \bibnamefont
  {Branco}}, \bibinfo {author} {\bibfnamefont {P.~M.}\ \bibnamefont
  {Ferreira}}, \bibinfo {author} {\bibfnamefont {L.}~\bibnamefont {Lavoura}},
  \bibinfo {author} {\bibfnamefont {M.~N.}\ \bibnamefont {Rebelo}}, \bibinfo
  {author} {\bibfnamefont {Marc}\ \bibnamefont {Sher}}, \ and\ \bibinfo
  {author} {\bibfnamefont {Joao~P.}\ \bibnamefont {Silva}},\ }\bibfield
  {title} {\enquote {\bibinfo {title} {{Theory and phenomenology of
  two-Higgs-doublet models}},}\ }\href {\doibase 10.1016/j.physrep.2012.02.002}
  {\bibfield  {journal} {\bibinfo  {journal} {Phys. Rept.}\ }\textbf {\bibinfo
  {volume} {516}},\ \bibinfo {pages} {1--102} (\bibinfo {year} {2012})},\
  \Eprint {http://arxiv.org/abs/1106.0034} {arXiv:1106.0034 [hep-ph]}
  \BibitemShut {NoStop}%
\bibitem [{\citenamefont {Cheng}\ and\ \citenamefont {Li}(1980)}]{Cheng1980}%
  \BibitemOpen
  \bibfield  {author} {\bibinfo {author} {\bibfnamefont {T.~P.}\ \bibnamefont
  {Cheng}}\ and\ \bibinfo {author} {\bibfnamefont {Ling-Fong}\ \bibnamefont
  {Li}},\ }\bibfield  {title} {\enquote {\bibinfo {title} {{Neutrino Masses,
  Mixings and Oscillations in SU(2) x U(1) Models of Electroweak
  Interactions}},}\ }\href {\doibase 10.1103/PhysRevD.22.2860} {\bibfield
  {journal} {\bibinfo  {journal} {Phys. Rev.}\ }\textbf {\bibinfo {volume}
  {D22}},\ \bibinfo {pages} {2860} (\bibinfo {year} {1980})}\BibitemShut
  {NoStop}%
\bibitem [{\citenamefont {Zee}(1985)}]{Zee1985}%
  \BibitemOpen
  \bibfield  {author} {\bibinfo {author} {\bibfnamefont {A.}~\bibnamefont
  {Zee}},\ }\bibfield  {title} {\enquote {\bibinfo {title} {Charged scalar
  field and quantum number violations},}\ }\href {\doibase
  https://doi.org/10.1016/0370-2693(85)90625-2} {\bibfield  {journal} {\bibinfo
   {journal} {Physics Letters B}\ }\textbf {\bibinfo {volume} {161}},\ \bibinfo
  {pages} {141 -- 145} (\bibinfo {year} {1985})}\BibitemShut {NoStop}%
\bibitem [{\citenamefont {Herrero-Garcia}\ \emph {et~al.}(2016)\citenamefont
  {Herrero-Garcia}, \citenamefont {Nebot}, \citenamefont {Rius},\ and\
  \citenamefont {Santamaria}}]{Herrero-garcia2014a}%
  \BibitemOpen
  \bibfield  {author} {\bibinfo {author} {\bibfnamefont {Juan}\ \bibnamefont
  {Herrero-Garcia}}, \bibinfo {author} {\bibfnamefont {Miguel}\ \bibnamefont
  {Nebot}}, \bibinfo {author} {\bibfnamefont {Nuria}\ \bibnamefont {Rius}}, \
  and\ \bibinfo {author} {\bibfnamefont {Arcadi}\ \bibnamefont {Santamaria}},\
  }\bibfield  {title} {\enquote {\bibinfo {title} {{Testing the Zee-Babu model
  via neutrino data, lepton flavour violation and direct searches at the
  LHC}},}\ }\bibfield  {booktitle} {\emph {\bibinfo {booktitle} {{Proceedings,
  37th International Conference on High Energy Physics (ICHEP 2014): Valencia,
  Spain, July 2-9, 2014}}},\ }\href {\doibase
  10.1016/j.nuclphysbps.2015.09.271} {\bibfield  {journal} {\bibinfo  {journal}
  {Nucl. Part. Phys. Proc.}\ }\textbf {\bibinfo {volume} {273-275}},\ \bibinfo
  {pages} {1678--1684} (\bibinfo {year} {2016})},\ \Eprint
  {http://arxiv.org/abs/1410.2299} {arXiv:1410.2299 [hep-ph]} \BibitemShut
  {NoStop}%
\bibitem [{\citenamefont {Fileviez~Perez}\ \emph {et~al.}(2008)\citenamefont
  {Fileviez~Perez}, \citenamefont {Han}, \citenamefont {Huang}, \citenamefont
  {Li},\ and\ \citenamefont {Wang}}]{FileviezPerez2008}%
  \BibitemOpen
  \bibfield  {author} {\bibinfo {author} {\bibfnamefont {Pavel}\ \bibnamefont
  {Fileviez~Perez}}, \bibinfo {author} {\bibfnamefont {Tao}\ \bibnamefont
  {Han}}, \bibinfo {author} {\bibfnamefont {Gui-yu}\ \bibnamefont {Huang}},
  \bibinfo {author} {\bibfnamefont {Tong}\ \bibnamefont {Li}}, \ and\ \bibinfo
  {author} {\bibfnamefont {Kai}\ \bibnamefont {Wang}},\ }\bibfield  {title}
  {\enquote {\bibinfo {title} {{Neutrino Masses and the CERN LHC: Testing Type
  II Seesaw}},}\ }\href {\doibase 10.1103/PhysRevD.78.015018} {\bibfield
  {journal} {\bibinfo  {journal} {Phys. Rev.}\ }\textbf {\bibinfo {volume}
  {D78}},\ \bibinfo {pages} {015018} (\bibinfo {year} {2008})},\ \Eprint
  {http://arxiv.org/abs/0805.3536} {arXiv:0805.3536 [hep-ph]} \BibitemShut
  {NoStop}%
\bibitem [{\citenamefont {Sugiyama}(2013)}]{Sugiyama2013}%
  \BibitemOpen
  \bibfield  {author} {\bibinfo {author} {\bibfnamefont {Hiroaki}\ \bibnamefont
  {Sugiyama}},\ }\bibfield  {title} {\enquote {\bibinfo {title} {{Neutrino Mass
  in TeV-Scale New Physics Models}},}\ }in\ \href
  {http://www.slac.stanford.edu/econf/C130213.1/pdfs/sugiyama.pdf} {\emph
  {\bibinfo {booktitle} {{Proceedings, 1st Toyama International Workshop on
  Higgs as a Probe of New Physics 2013 (HPNP2013): Toyama, Japan, February
  13-16, 2013}}}}\ (\bibinfo {year} {2013})\ \Eprint
  {http://arxiv.org/abs/1304.6031} {arXiv:1304.6031 [hep-ph]} \BibitemShut
  {NoStop}%
\bibitem [{\citenamefont {Das}\ and\ \citenamefont
  {Santamaria}(2016)}]{Das2016}%
  \BibitemOpen
  \bibfield  {author} {\bibinfo {author} {\bibfnamefont {Dipankar}\
  \bibnamefont {Das}}\ and\ \bibinfo {author} {\bibfnamefont {Arcadi}\
  \bibnamefont {Santamaria}},\ }\bibfield  {title} {\enquote {\bibinfo {title}
  {{Updated scalar sector constraints in the Higgs triplet model}},}\ }\href
  {\doibase 10.1103/PhysRevD.94.015015} {\bibfield  {journal} {\bibinfo
  {journal} {Phys. Rev.}\ }\textbf {\bibinfo {volume} {D94}},\ \bibinfo {pages}
  {015015} (\bibinfo {year} {2016})},\ \Eprint
  {http://arxiv.org/abs/1604.08099} {arXiv:1604.08099 [hep-ph]} \BibitemShut
  {NoStop}%
\bibitem [{\citenamefont {Montero}\ \emph {et~al.}(1999)\citenamefont
  {Montero}, \citenamefont {de~Sousa~Pires},\ and\ \citenamefont
  {Pleitez}}]{Montero1999}%
  \BibitemOpen
  \bibfield  {author} {\bibinfo {author} {\bibfnamefont {J.~C.}\ \bibnamefont
  {Montero}}, \bibinfo {author} {\bibfnamefont {Carlos~Antonio}\ \bibnamefont
  {de~Sousa~Pires}}, \ and\ \bibinfo {author} {\bibfnamefont {V.}~\bibnamefont
  {Pleitez}},\ }\bibfield  {title} {\enquote {\bibinfo {title} {{Spontaneous
  breaking of a global symmetry in a 331 model}},}\ }\href {\doibase
  10.1103/PhysRevD.60.115003} {\bibfield  {journal} {\bibinfo  {journal} {Phys.
  Rev.}\ }\textbf {\bibinfo {volume} {D60}},\ \bibinfo {pages} {115003}
  (\bibinfo {year} {1999})},\ \Eprint {http://arxiv.org/abs/hep-ph/9903251}
  {arXiv:hep-ph/9903251 [hep-ph]} \BibitemShut {NoStop}%
\bibitem [{\citenamefont {Langacker}(2009)}]{Langacker:1226768}%
  \BibitemOpen
  \bibfield  {author} {\bibinfo {author} {\bibfnamefont {P.}~\bibnamefont
  {Langacker}},\ }\href {https://books.google.ca/books?id=dpANo3e\_pS8C} {\emph
  {\bibinfo {title} {The Standard Model and Beyond}}},\ Series in High Energy
  Physics, Cosmology and Gravitation\ (\bibinfo  {publisher} {CRC Press},\
  \bibinfo {year} {2009})\BibitemShut {NoStop}%
\bibitem [{\citenamefont {Wu}\ and\ \citenamefont {Zhou}(2001)}]{Wu2001}%
  \BibitemOpen
  \bibfield  {author} {\bibinfo {author} {\bibfnamefont {Yue-Liang}\
  \bibnamefont {Wu}}\ and\ \bibinfo {author} {\bibfnamefont {Yu-Feng}\
  \bibnamefont {Zhou}},\ }\bibfield  {title} {\enquote {\bibinfo {title} {{Muon
  anomalous magnetic moment in the standard model with two Higgs doublets}},}\
  }\href {\doibase 10.1103/PhysRevD.64.115018} {\bibfield  {journal} {\bibinfo
  {journal} {Phys. Rev.}\ }\textbf {\bibinfo {volume} {D64}},\ \bibinfo {pages}
  {115018} (\bibinfo {year} {2001})},\ \Eprint
  {http://arxiv.org/abs/hep-ph/0104056} {arXiv:hep-ph/0104056 [hep-ph]}
  \BibitemShut {NoStop}%
\bibitem [{\citenamefont {Dicus}\ \emph {et~al.}(2001)\citenamefont {Dicus},
  \citenamefont {He},\ and\ \citenamefont {Ng}}]{Dicus2001}%
  \BibitemOpen
  \bibfield  {author} {\bibinfo {author} {\bibfnamefont {Duane~A.}\
  \bibnamefont {Dicus}}, \bibinfo {author} {\bibfnamefont {Hong-Jian}\
  \bibnamefont {He}}, \ and\ \bibinfo {author} {\bibfnamefont {John~N.}\
  \bibnamefont {Ng}},\ }\bibfield  {title} {\enquote {\bibinfo {title}
  {{Neutrino - lepton masses, Zee scalars and muon g-2}},}\ }\href {\doibase
  10.1103/PhysRevLett.87.111803} {\bibfield  {journal} {\bibinfo  {journal}
  {Phys. Rev. Lett.}\ }\textbf {\bibinfo {volume} {87}},\ \bibinfo {pages}
  {111803} (\bibinfo {year} {2001})},\ \Eprint
  {http://arxiv.org/abs/hep-ph/0103126} {arXiv:hep-ph/0103126 [hep-ph]}
  \BibitemShut {NoStop}%
\bibitem [{\citenamefont {de~S.~Pires}\ and\ \citenamefont {Rodrigues~da
  Silva}(2001)}]{Pires2001}%
  \BibitemOpen
  \bibfield  {author} {\bibinfo {author} {\bibfnamefont {C.~A.}\ \bibnamefont
  {de~S.~Pires}}\ and\ \bibinfo {author} {\bibfnamefont {P.~S.}\ \bibnamefont
  {Rodrigues~da Silva}},\ }\bibfield  {title} {\enquote {\bibinfo {title}
  {{Scalar scenarios contributing to (g-2)(muon) with enhanced Yukawa
  couplings}},}\ }\href {\doibase 10.1103/PhysRevD.64.117701} {\bibfield
  {journal} {\bibinfo  {journal} {Phys. Rev.}\ }\textbf {\bibinfo {volume}
  {D64}},\ \bibinfo {pages} {117701} (\bibinfo {year} {2001})},\ \Eprint
  {http://arxiv.org/abs/hep-ph/0103083} {arXiv:hep-ph/0103083 [hep-ph]}
  \BibitemShut {NoStop}%
\bibitem [{\citenamefont {Pich}(2014)}]{Pich2014}%
  \BibitemOpen
  \bibfield  {author} {\bibinfo {author} {\bibfnamefont {Antonio}\ \bibnamefont
  {Pich}},\ }\bibfield  {title} {\enquote {\bibinfo {title} {{Precision Tau
  Physics}},}\ }\href {\doibase 10.1016/j.ppnp.2013.11.002} {\bibfield
  {journal} {\bibinfo  {journal} {Prog. Part. Nucl. Phys.}\ }\textbf {\bibinfo
  {volume} {75}},\ \bibinfo {pages} {41--85} (\bibinfo {year} {2014})},\
  \Eprint {http://arxiv.org/abs/1310.7922} {arXiv:1310.7922 [hep-ph]}
  \BibitemShut {NoStop}%
\bibitem [{\citenamefont {Baldini}\ \emph {et~al.}(2016)\citenamefont {Baldini}
  \emph {et~al.}}]{TheMEG:2016wtm}%
  \BibitemOpen
  \bibfield  {author} {\bibinfo {author} {\bibfnamefont {A.~M.}\ \bibnamefont
  {Baldini}} \emph {et~al.} (\bibinfo {collaboration} {MEG}),\ }\bibfield
  {title} {\enquote {\bibinfo {title} {{Search for the lepton flavour violating
  decay $\mu ^+ \rightarrow \mathrm {e}^+ \gamma $ with the full dataset of the
  MEG experiment}},}\ }\href {\doibase 10.1140/epjc/s10052-016-4271-x}
  {\bibfield  {journal} {\bibinfo  {journal} {Eur. Phys. J.}\ }\textbf
  {\bibinfo {volume} {C76}},\ \bibinfo {pages} {434} (\bibinfo {year}
  {2016})},\ \Eprint {http://arxiv.org/abs/1605.05081} {arXiv:1605.05081
  [hep-ex]} \BibitemShut {NoStop}%
\bibitem [{\citenamefont {Merle}\ and\ \citenamefont
  {Rodejohann}(2006)}]{Merle2006}%
  \BibitemOpen
  \bibfield  {author} {\bibinfo {author} {\bibfnamefont {Alexander}\
  \bibnamefont {Merle}}\ and\ \bibinfo {author} {\bibfnamefont {Werner}\
  \bibnamefont {Rodejohann}},\ }\bibfield  {title} {\enquote {\bibinfo {title}
  {{The Elements of the neutrino mass matrix: Allowed ranges and implications
  of texture zeros}},}\ }\href {\doibase 10.1103/PhysRevD.73.073012} {\bibfield
   {journal} {\bibinfo  {journal} {Phys. Rev.}\ }\textbf {\bibinfo {volume}
  {D73}},\ \bibinfo {pages} {073012} (\bibinfo {year} {2006})},\ \Eprint
  {http://arxiv.org/abs/hep-ph/0603111} {arXiv:hep-ph/0603111 [hep-ph]}
  \BibitemShut {NoStop}%
\bibitem [{\citenamefont {Grimus}\ and\ \citenamefont
  {Ludl}(2012)}]{Grimus2012}%
  \BibitemOpen
  \bibfield  {author} {\bibinfo {author} {\bibfnamefont {W.}~\bibnamefont
  {Grimus}}\ and\ \bibinfo {author} {\bibfnamefont {P.~O.}\ \bibnamefont
  {Ludl}},\ }\bibfield  {title} {\enquote {\bibinfo {title} {{Correlations of
  the elements of the neutrino mass matrix}},}\ }\href {\doibase
  10.1007/JHEP12(2012)117} {\bibfield  {journal} {\bibinfo  {journal} {JHEP}\
  }\textbf {\bibinfo {volume} {12}},\ \bibinfo {pages} {117} (\bibinfo {year}
  {2012})},\ \Eprint {http://arxiv.org/abs/1209.2601} {arXiv:1209.2601
  [hep-ph]} \BibitemShut {NoStop}%
\bibitem [{\citenamefont {Chatrchyan}\ \emph {et~al.}(2012)\citenamefont
  {Chatrchyan} \emph {et~al.}}]{Chatrchyan:2012ya}%
  \BibitemOpen
  \bibfield  {author} {\bibinfo {author} {\bibfnamefont {Serguei}\ \bibnamefont
  {Chatrchyan}} \emph {et~al.} (\bibinfo {collaboration} {CMS}),\ }\bibfield
  {title} {\enquote {\bibinfo {title} {{A search for a doubly-charged Higgs
  boson in $pp$ collisions at $\sqrt{s}=7$ TeV}},}\ }\href {\doibase
  10.1140/epjc/s10052-012-2189-5} {\bibfield  {journal} {\bibinfo  {journal}
  {Eur. Phys. J.}\ }\textbf {\bibinfo {volume} {C72}},\ \bibinfo {pages} {2189}
  (\bibinfo {year} {2012})},\ \Eprint {http://arxiv.org/abs/1207.2666}
  {arXiv:1207.2666 [hep-ex]} \BibitemShut {NoStop}%
\bibitem [{\citenamefont {Aad}\ \emph {et~al.}(2012)\citenamefont {Aad} \emph
  {et~al.}}]{ATLAS:2012hi}%
  \BibitemOpen
  \bibfield  {author} {\bibinfo {author} {\bibfnamefont {Georges}\ \bibnamefont
  {Aad}} \emph {et~al.} (\bibinfo {collaboration} {ATLAS}),\ }\bibfield
  {title} {\enquote {\bibinfo {title} {{Search for doubly-charged Higgs bosons
  in like-sign dilepton final states at $\sqrt{s}=7$ TeV with the ATLAS
  detector}},}\ }\href {\doibase 10.1140/epjc/s10052-012-2244-2} {\bibfield
  {journal} {\bibinfo  {journal} {Eur. Phys. J.}\ }\textbf {\bibinfo {volume}
  {C72}},\ \bibinfo {pages} {2244} (\bibinfo {year} {2012})},\ \Eprint
  {http://arxiv.org/abs/1210.5070} {arXiv:1210.5070 [hep-ex]} \BibitemShut
  {NoStop}%
\bibitem [{\citenamefont {del Aguila}\ and\ \citenamefont
  {Chala}(2014)}]{delAguila:2013mia}%
  \BibitemOpen
  \bibfield  {author} {\bibinfo {author} {\bibfnamefont {Francisco}\
  \bibnamefont {del Aguila}}\ and\ \bibinfo {author} {\bibfnamefont {Mikael}\
  \bibnamefont {Chala}},\ }\bibfield  {title} {\enquote {\bibinfo {title} {{LHC
  bounds on Lepton Number Violation mediated by doubly and singly-charged
  scalars}},}\ }\href {\doibase 10.1007/JHEP03(2014)027} {\bibfield  {journal}
  {\bibinfo  {journal} {JHEP}\ }\textbf {\bibinfo {volume} {03}},\ \bibinfo
  {pages} {027} (\bibinfo {year} {2014})},\ \Eprint
  {http://arxiv.org/abs/1311.1510} {arXiv:1311.1510 [hep-ph]} \BibitemShut
  {NoStop}%
\bibitem [{\citenamefont {Couchot}\ \emph {et~al.}(2017)\citenamefont
  {Couchot}, \citenamefont {Henrot-Versill\'e}, \citenamefont {Perdereau},
  \citenamefont {Plaszczynski}, \citenamefont {d'Orfeuil}, \citenamefont
  {Spinelli},\ and\ \citenamefont {Tristram}}]{Couchot2017}%
  \BibitemOpen
  \bibfield  {author} {\bibinfo {author} {\bibfnamefont {F.}~\bibnamefont
  {Couchot}}, \bibinfo {author} {\bibfnamefont {S.}~\bibnamefont
  {Henrot-Versill\'e}}, \bibinfo {author} {\bibfnamefont {O.}~\bibnamefont
  {Perdereau}}, \bibinfo {author} {\bibfnamefont {S.}~\bibnamefont
  {Plaszczynski}}, \bibinfo {author} {\bibfnamefont {B.~Rouill\'e}\
  \bibnamefont {d'Orfeuil}}, \bibinfo {author} {\bibfnamefont {M.}~\bibnamefont
  {Spinelli}}, \ and\ \bibinfo {author} {\bibfnamefont {M.}~\bibnamefont
  {Tristram}},\ }\href@noop {} {\enquote {\bibinfo {title} {{Cosmological
  constraints on the neutrino mass including systematic uncertainties}},}\ }
  (\bibinfo {year} {2017}),\ \Eprint {http://arxiv.org/abs/1703.10829}
  {arXiv:1703.10829 [astro-ph.CO]} \BibitemShut {NoStop}%
\bibitem [{\citenamefont {Olive}\ \emph {et~al.}(2014)\citenamefont {Olive}
  \emph {et~al.}}]{Agashe2014}%
  \BibitemOpen
  \bibfield  {author} {\bibinfo {author} {\bibfnamefont {K.~A.}\ \bibnamefont
  {Olive}} \emph {et~al.} (\bibinfo {collaboration} {Particle Data Group}),\
  }\bibfield  {title} {\enquote {\bibinfo {title} {{Review of Particle
  Physics}},}\ }\href {\doibase 10.1088/1674-1137/38/9/090001} {\bibfield
  {journal} {\bibinfo  {journal} {Chin. Phys.}\ }\textbf {\bibinfo {volume}
  {C38}},\ \bibinfo {pages} {090001} (\bibinfo {year} {2014})}\BibitemShut
  {NoStop}%
\end{thebibliography}%


\end{document}